\documentclass[12pt]{article}

\usepackage{amsmath,amsfonts,amssymb}
\usepackage{booktabs}
\usepackage{array}
\usepackage{mathptmx}
\usepackage{txfonts}
\usepackage[normalem]{ulem}
\usepackage{graphicx}
\usepackage{tikz}
\usepackage{subcaption}
\newcolumntype{L}[1]{>{\raggedright\arraybackslash}p{#1}}
\usepackage{booktabs,tabularx,array,makecell,multirow,siunitx,pifont}
\newcolumntype{C}{>{\centering\arraybackslash}X}
\usetikzlibrary{arrows.meta, positioning, shapes.geometric}
\usepackage{amsmath}
\usepackage{amssymb}
\usepackage{bm}
\tikzstyle{block} = [rectangle, draw=black, minimum width=4cm, minimum height=1.2cm, text centered]
\tikzstyle{parallelogram} = [trapezium, trapezium left angle=75, trapezium right angle=105, draw=black, minimum height=1cm, text centered]
\tikzstyle{arrow} = [thick, -{Latex[length=2mm]}]
\newcolumntype{L}{>{\raggedright\arraybackslash}X}
\newcommand{\dm}[1]{\makecell[l]{#1}} 
\newcommand{\bn}{\boldsymbol{n}}

\usepackage{graphicx}

\usepackage[letterpaper,margin=1in]{geometry}

\usepackage[nolist,nohyperlinks]{acronym} %
\usepackage[colorlinks=true,unicode,linkcolor=blue,citecolor=cyan,]{hyperref}
\usepackage{esvect}

\renewenvironment{abstract}
	{\quotation}
	{\endquotation}

\date{}

\makeatletter
\renewcommand{\fnum@figure}{\textbf{Figure \thefigure}}
\renewcommand{\fnum@table}{\textbf{Table \thetable}}
\makeatother

\usepackage{scicite}

\usepackage{url}

\def\scititle{
Energetic Origins of Competing Deformation Modes in Metastable Titanium Alloys
}
\title{\bfseries \boldmath \scititle}

\author{
        Ganlin Chen$^{1}$,
        Deepak V Pillai$^{2}$,
        Yufeng Zheng$^{2}$,
        Liang~Qi$^{1,\ast}$\and
	\small$^{1}$Department of Materials Science and Engineering, University of Michigan, Ann Arbor, MI 48109, USA.\and
    \small$^{2}$Department of Materials Science and Engineering, University of North Texas, Denton, TX 76207, USA.\and
	\small$^\ast$Corresponding author. Email: qiliang@umich.edu
    }

\begin{document} 
\begin{acronym}
    \acro{DFT}{density function theory}
    \acro{MD}{molecular dynamics}
    \acro{MBE}{molecular-beam epitaxy}
    \acro{MEP}{minimum energy path}
    \acro{STEM}{Scanning transmission electron microscopy}
    \acro{TEM}{transmission electron microscopy}
    \acro{VASP}{Vienna \textit{Ab Initio} Simulation Package}
    \acro{MEAM}{modified embedded atom method}
\end{acronym}

\maketitle
\begin{abstract} 
\bfseries \boldmath
Metastable alloys, such as $\beta$-phase titanium (Ti) alloys with a body-centered cubic (BCC) lattice, can exhibit exceptional mechanical properties through the interplay of multiple deformation mechanisms—diffusionless phase transformations, deformation twinning, and conventional dislocation slip. However, understanding how these mechanisms compete or cooperate across a wide range of metastable alloys and loading conditions remains a fundamental challenge. Here, we employ molecular dynamics (MD) simulations to investigate the nucleation behavior of competing deformation modes in metastable $\beta$-Ti alloys as a function of temperature, composition, and loading conditions. We reveal that twinning pathways emerge through reversible transformations between the $\beta$ phase and the orthorhombic $\alpha^{\prime\prime}$ phase, in agreement with crystallographic theories. Quantitative analyses demonstrate that the dominant deformation mechanisms and preferred twinning-plane orientations are governed by two key energetic parameters: the free energy barrier for homogeneous $\beta \leftrightarrow \alpha^{\prime\prime}$ transformations and the misfit strain energy along specific phase boundaries. These energetic quantities vary systematically with thermodynamic and mechanical conditions, thereby rationalizing the deformation mode transitions observed in both simulations and experiments. These energetic metrics offer a physically grounded and computationally tractable basis for designing next-generation metastable alloys.

\end{abstract}

\section*{Teaser}
An experimentally measurable theoretical framework unveils deformation modes selection in meta-stable Titanium alloys.

\section*{Introduction}
\label{intro}
The increasing demand from energy, transportation, biomedical, and aerospace sectors for structural materials that simultaneously exhibit high strength, excellent ductility, strong work-hardening capacity, and reliable performance under extreme conditions has intensified efforts to design meta-stable alloys~\cite{raabe2019metastability}. Such alloys enable the activation of multiple deformation mechanisms to achieve superior mechanical properties. Among them, metastable $\beta$-titanium ($\beta$-Ti) alloys are particularly attractive owing to their low density, outstanding corrosion and high-temperature resistance, and excellent biocompatibility~\cite{kolli2018review,pesode2023review}. In these $\beta$-Ti alloys, the interplay among twinning-induced plasticity (TWIP) from deformation twinning, transformation-induced plasticity (TRIP) arising from diffusionless phase transformations, and dislocation slip plays a critical role in optimizing mechanical properties~\cite{khachaturyan2013theory,christian1975theory,christian1995deformation,saito2003multifunctional,bouaziz2011high,de2018twinning,fischer1996transformation,fischer2000new,olson1978transformation,banerjee2013perspectives,boyer1996overview,brozek2016beta,leyens2006titanium,niinomi2008mechanical,marteleur2012design,hao2016superelasticity,li2023deformation}. Comparable complex deformation behaviors are also observed in metastable BCC refractory high-entropy alloys (HEAs) that exhibit exceptional mechanical performance~\cite{wang2024tizrhfnb,lilensten2017design,huang2017phase,wang2019formation,zhang2018phase,wang2020mechanical}. Achieving superior properties requires activating these mechanisms in the proper sequence and to the appropriate extent, which demands precise engineering of their thermodynamic and kinetic characteristics as well as processing pathways~\cite{cao2024review,gao2018deformation,hanada1987correlation,min2013mechanism,sun2013investigation,blackburn1971stress,zhang2019strong}. Despite decades of research, however, a comprehensive and robust strategy for guiding the controllable activation of these mechanisms in $\beta$-Ti and similar alloy systems has yet to be fully established.

A major challenge in accurately predicting deformation mode activation and their interactions in $\beta$-Ti alloys arises from the coexistence and strong coupling of multiple metastable phase transformations and deformation twinning modes. The principal metastable phases in Ti alloys include the $\beta$ phase with a BCC lattice, the $\alpha'$ phase with a hexagonal close-packed (HCP) structure, the $\alpha''$ phase with an orthorhombic lattice, and the $\omega$ phase with a hexagonal lattice~\cite{gao2016group,gao2020intrinsic}. Among deformation twins, the conventional \{112\}$\langle$11$\bar{1}$$\rangle$ twinning system is common in BCC transition metals and alloys~\cite{groger2023twinning}. However, in $\beta$-Ti alloys, high-index twinning systems (Miller indices $>2$) are frequently observed~\cite{zhang2019strong,antonov2020novel}, most notably the \{332\}$\langle$11$\bar{3}$$\rangle$ twin as the dominant twinning behavior~\cite{crocker1962twinned,tobe2014origin,hanada1985deformation,zhang2020hierarchical,chen2018transitional,lai2016mechanism,oka1978stress,oka1979332,blackburn1971stress,hanada1986transmission}. The \{332\} twins frequently coexist with \{112\} twins, producing complex hierarchical microstructures. For example, Figure~\ref{fig:Coexist} (A–C) shows TEM images of \{112\} and \{332\} twins in a Ti–24Nb–4Zr–8Sn (Ti2448) alloy after 5\% cold rolling, with schematic illustrations of the corresponding twin orientations in Figure~\ref{fig:Coexist} (D–E). Crystallographic theory further predicts a strong interplay between reversible diffusionless phase transformations and deformation twinning~\cite{gao2020intrinsic,bhattacharya2003microstructure}. As illustrated in Figure~\ref{fig:Coexist} (F), one possible pathway for twin formation involves a reversible $\beta_{\textup{matrix}}$ $\rightarrow$ $\alpha''$ $\rightarrow$ $\beta_{\textup{twin}}$ transformation~\cite{gao2020intrinsic}, wherein both \{332\} and \{112\} twins emerge as conjugate solutions of the kinematic compatibility condition between the $\beta$ matrix and the twin variant. The coupling between transformations and twinning is further supported by experimental and computational observations of $\omega$ and $\alpha''$ phases at twin boundaries~\cite{wu20141,chen2024stability,lai2016mechanism,cho2020study,takemoto1993structural,zheng2016effect,chen2018transitional,castany2016reversion}.

The coupling between phase transformations and twinning suggests that the preferred deformation modes can be correlated with, and even predicted by, the phase stability of metastable Ti alloys~\cite{hanada1986effect,zhao2021materials}. For instance, in Ti–Nb alloys, the \{332\} twin dominates when BCC stability is low (low Nb concentration). As BCC stability increases slightly, its formation becomes temperature- and orientation-dependent, with a transition to \{112\} twinning and eventually to dislocation slip~\cite{hanada1985deformation}. Similar correlations between deformation modes and BCC stability have also been observed in Ti–V and Ti–Mo alloys~\cite{hanada1986effect}. To rationalize these trends, the empirical $\overline{\text{Bo}}$–$\overline{\text{Md}}$ diagram, inspired by Hume–Rothery rules~\cite{mizutani2010hume}, relates deformation mode variations to the mean electronic parameters $\overline{\text{Bo}}$ (bond order) and $\overline{\text{Md}}$ (metal $d$-orbital energy level), and has successfully explained general trends in Ti alloys~\cite{morinaga1988theoretical}. However, the $\overline{\text{Bo}}$–$\overline{\text{Md}}$ diagram breaks down in more complex alloy systems~\cite{coffigniez2024combination} and provides no guidance on temperature effects or processing conditions. A more recent model based on first-principles calculations of equilibrium elastic properties at 0~K also fails to capture temperature effects, twinning competitions, or dynamic deformation behavior under load~\cite{lv2025novel}. Similarly, a physics-based model that estimates free-energy barriers for phase transformation nucleation under external force predicts preferred deformation modes (TRIP, TWIP, or dislocation) but cannot distinguish between different twinning modes~\cite{zhao2021materials}. A key shortcoming of these approaches is the lack of accurate consideration of free energy variations along the full twinning pathway. Gao \textit{et al.}~\cite{gao2020intrinsic} proposed a rigorous framework in which twin formation occurs via multi-step reversible phase transformations, suggesting a promising route to construct energetic models along these pathways for more accurate predictions of TWIP and TRIP behavior.



To elucidate the formation mechanisms and energetics of different deformation modes, particularly the complete twinning pathway, we employ \ac{MD} simulations to systematically probe the nucleation of deformation modes in representative Ti–Nb and Ti–Mo alloys. The simulations quantitatively reveal how temperature, chemical composition, and loading conditions govern the competition among different deformation modes. We demonstrate that both \{112\} and \{332\} twinning pathways emerge via reversible transformations between the $\beta$ and $\alpha''$ phases, consistent with crystallographic models~\cite{gao2020intrinsic}. Quantitative analyses further show that the dominant deformation modes and preferred twinning-plane orientations (\{112\} vs.\ \{332\}) are dictated by both the chemical free-energy changes along the $\beta_{\textup{matrix}} \rightarrow \alpha'' \rightarrow \beta_{\textup{twin}}$ transformation pathway and by the misfit strain energies between these phases along the corresponding twinning-plane orientations. These energetic parameters vary systematically with temperature, composition, and loading method, thereby explaining the deformation-mode transitions observed in both simulations and experiments. Notably, the contributions of misfit strain energies to the twinning nucleation barrier are generally comparable to those of chemical free-energy changes, indicating that models based solely on thermodynamic phase stability and related parameters are insufficient to capture all key factors governing twinning-mode selection~\cite{morinaga1988theoretical}. Finally, we propose a conceptual flowchart based on these energetic metrics to guide the design of Ti alloys with tailored mechanical performance. This framework, grounded in these two critical energetic parameters, is extendable to metastable Ti alloys and refractory HEAs and adaptable for machine-learning–assisted high-throughput computational screening, thereby accelerating alloy discovery.

\section*{Results}
\label{result}

\paragraph*{Molecular dynamics (MD) simulations on deformation modes in $\beta$ Ti-Nb alloys}\label{para:MD}

\label{MD_Structural}
We first examine the atomistic-scale nucleation behaviors of common deformation modes in $\beta$-Ti alloys as functions of temperature, composition, and loading conditions. These modes include stress-induced diffusionless transformations (primarily $\alpha''$), \{112\} twinning, \{332\} twinning, and dislocation slip. Among them, the nucleation of \{332\} twins is particularly complex, and few atomistic studies have reproduced it in a manner consistent with experimental observations. To address this, we apply shear loading along the [311] direction on $(\bar{2}33)$ planes, which provides the most favorable driving force for \{332\} twin formation~\cite{tobe2014origin}. Since alternative mechanisms may still dominate under certain conditions~\cite{hanada1985deformation}, we perform MD simulations across a range of alloy compositions, temperatures, and boundary conditions to investigate deformation mode selection under different thermomechanical environments. Specifically, two representative temperatures are considered: 77~K (liquid nitrogen temperature) and 300~K, both widely used in experiments. For compositional variations, we select Ti$_{5}$Nb (close to the Nb concentration in the experimental sample shown in Figure~\ref{fig:Coexist}), Ti$_{4}$Nb, and Ti$_{2}$Nb, representing alloys spanning distinct regimes of $\beta$-phase stability. For mechanical boundary conditions under shear loading, we employ two setups: ``Non-Confined Simple Shear’’ (NSS) and ``Confined Simple Shear’’ (CSS). In the NSS case, shear strain is incrementally applied while keeping all normal and non-applied shear stresses zero, simulating pure shear deformation under stress-free conditions. In the CSS case, shear strain is incrementally applied while keeping all non-shear strain components fixed at zero, simulating mechanically constrained environments such as local confined grains in bulk samples. Further details of the MD settings are provided in the \textbf{Materials and Methods}.

\begin{figure}[htbp]
    \centering
    \makebox[\textwidth][c]{%
    \includegraphics[width=0.8\textwidth, clip=true, trim=0 0 0 0]{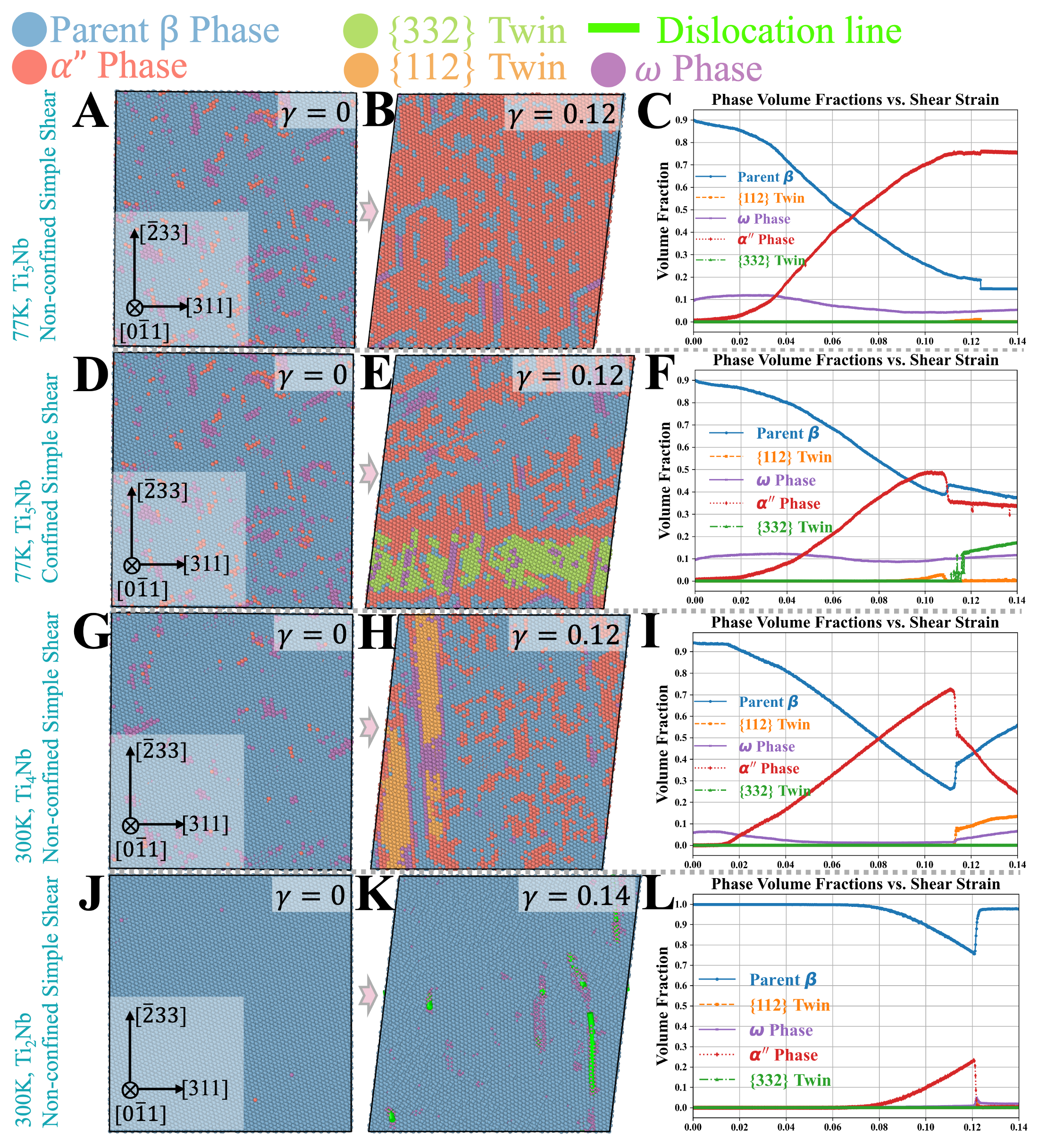}
    }
    \caption{\textbf{Atomistic structural evolution and deformation modes of Ti-Nb alloys under varying temperature, composition, and stress/strain conditions by MD simulations. In all cases, the shear loading is along [311] direction on ($\bar{2}$33) planes.} Temperature, composition and stress/strain conditions are fixed in each row: \textbf{(A-C)} is Ti$_{5}$Nb under 77~K, non-confined simple shear condition; \textbf{(D-F)} is Ti$_{5}$Nb under 77~K, confined simple shear condition; \textbf{(G-I)} is Ti$_{4}$Nb under 300~K, non-confined simple shear condition; \textbf{(J-L)} is Ti$_{2}$Nb under 300~K, non-confined simple shear condition. From top to down, \textbf{(A,D,G,J)} are snapshots of atomistic structures before loading applied; \textbf{(B,E,H,K)} are snapshots of atomistic structures when applied shear strain is 0.12 and different deformation modes produced; \textbf{(C,F,I,L)} illustrate nucleation of different metastable phases during shear loading and dynamic volume fraction change. Color interpretation of atoms in the MD snapshots locating at the top of the figure. dominant deformation mode of each row is different: \textbf{(A-C)} is $\alpha"$ martensite; \textbf{(D-F)} is \{332\} twin; \textbf{(G-I)} is \{112\} twin and \textbf{(J-L)} is dislocation slip.}
    \label{fig:Structural}
\end{figure}

Figure~\ref{fig:Structural} shows MD simulation results for the nucleation and early-stage growth of different deformation modes. The chemical composition, temperature, and loading conditions for each case are listed on the left of each row. Because $\beta$-phase stability in Ti–Nb alloys increases with both Nb content and temperature~\cite{ehemann2017force,chen2024stability}, we qualitatively rank the four cases in Figure~\ref{fig:Structural} by increasing stability: Ti$_5$Nb at 77~K (Figure~\ref{fig:Structural}(A–F)) exhibits the lowest $\beta$ stability, Ti$_4$Nb at 300~K (Figure~\ref{fig:Structural}(G–I)) shows intermediate stability, and Ti$_2$Nb at 300~K (Figure~\ref{fig:Structural}(J–L)) displays the highest stability. Each row corresponds to a specific alloy system and includes, from left to right: (i) the initial stress-free atomic configuration prior to shear loading; (ii) atomic snapshots under shear deformation; and (iii) the time evolution of the volume fraction of metastable phases and twinning regions. Atoms are color-coded to distinguish the $\beta$ phase (with different orientations due to twinning) from other metastable phases. Details of the phase-classification procedure and misorientation calculations are provided in the \textbf{Materials and Methods}.

When both the Nb concentration and temperature are relatively low, specifically for Ti$_5$Nb at 77~K, our MD simulations reveal the formation of locally distributed $\omega$ and $\alpha''$ phases (Figure~\ref{fig:Structural}(A,D)), consistent with typical diffusionless transformations in metastable $\beta$-Ti alloys~\cite{banerjee2013perspectives}. As mentioned above, two distinct shear loading conditions are examined: NSS (Figure~\ref{fig:Structural}(A--C)) and CSS (Figure~\ref{fig:Structural}(D--F)). Under NSS condition, a large fraction of the $\beta$ matrix (blue atoms) transforms into the $\alpha''$ phase (red atoms), with the phase fraction rapidly increasing and saturating at approximately 80\% when the applied shear strain reaches 0.12 (Figure~\ref{fig:Structural}(C)). Throughout this deformation, minimal formation of new $\beta$ twin variants is observed, indicating that the dominant mechanism is a direct $\beta\rightarrow\alpha''$ diffusionless transformation, which is retained under strain, exemplifying classical TRIP behavior widely reported in low-stability $\beta$-Ti alloys~\cite{banerjee2013perspectives}. Interestingly, when the loading condition is changed to CSS condition, the dominant deformation mechanism shifts. As shown in Figure~\ref{fig:Structural}(E), the nucleation and growth of $\alpha''$ are significantly suppressed, with the phase fraction plateauing at about 50\% at a shear strain of 0.11 (Figure~\ref{fig:Structural}(F)). Beyond this point, the $\alpha''$ fraction abruptly decreases by ~15\%, coinciding with the sudden appearance and growth of \{332\} twin structures, whose volume fraction increases from zero to ~15\% (green curve). Time-resolved atomic snapshots (Supplementary Movies) clearly show that \{332\} twin nucleation initiates within the $\alpha''$ phase. This twin formation mechanism—induced by reversible $\beta$-$\alpha''$ transformations—is supported by numerous experimental studies~\cite{lai2016mechanism,cho2020study,takemoto1993structural,zheng2016effect,chen2018transitional,castany2016reversion}. As shear strain continues to increase, the $\alpha''$ phase fraction remains relatively stable but localized primarily near \{332\} twin boundaries (Supplementary Materials). Simultaneously, the $\beta$ matrix fraction decreases, suggesting that \{332\} twins continue to grow at the expense of the parent phase.

With increasing $\beta$-phase stability, as in Ti$_{4}$Nb at 300~K (Figure~\ref{fig:Structural}(G–I)) and Ti$_{2}$Nb at 300~K (Figure~\ref{fig:Structural}(J–L)) under NSS, the dominant deformation mechanisms shift from $\beta \rightarrow \alpha''$ transformation and \{332\} twinning to \{112\} twinning and dislocation slip. In Ti$_{4}$Nb at 300~K, a twin domain (orange in Figure~\ref{fig:Structural}H) is identified as a \{112\} twin (see \textbf{Supplementary Materials}). Phase evolution tracking (Figure~\ref{fig:Structural}I) shows that, between strains of 0 and 0.11, the $\beta$ parent fraction (blue) decreases while $\alpha''$ (red) increases. Beyond this point, $\alpha''$ rapidly transforms into the \{112\} twin (orange), indicating a sequence of $\beta \rightarrow \alpha'' \rightarrow \beta \{112\} \textup{ twin}$. After nucleation, however, the \{112\} twin fraction remains nearly constant, while the $\beta$ matrix continues to grow. This suggests that although twin formation from $\alpha''$ is energetically favorable, subsequent twin growth is kinetically hindered, likely due to pinning by $\omega$ particles (purple in Figure~\ref{fig:Structural}H) observed in metastable $\beta$-Ti alloys~\cite{chen2024stability,wu20141}. In Ti$_{2}$Nb at 300~K, where $\beta$ stability is highest, phase transformations are largely suppressed (Figures~\ref{fig:Structural}(J,K)). The $\alpha''$ phase nucleates only at higher strains ($\varepsilon \approx 0.08$, Figure~\ref{fig:Structural}L), reaches a modest fraction ($\sim$20\%), and subsequently reverts to $\beta$ with the onset of dislocation slip, confirming that slip dominates deformation in highly stable $\beta$-Ti alloys.

Beyond the four representative cases in Figure~\ref{fig:Structural}, the dominant deformation modes across all simulated conditions are summarized in Table~\ref{tab:deformationMode}. These simulations span a wide range of thermodynamic $\beta$-phase stabilities, from the unstable/metastable boundary to the metastable/stable regime~\cite{ehemann2017force,chen2024stability}. At low stability, exemplified by Ti$_{5}$Nb at 77~K, deformation is dominated by $\beta \rightarrow \alpha''$ transformation and \{332\} twinning, with the prevailing mechanism depending on the loading condition (NSS vs.\ CSS). These behaviors are characteristic of TRIP and TWIP effects commonly observed in metastable $\beta$-Ti alloys. With moderate stability (e.g., Ti$_{4}$Nb at 77~K), $\beta \rightarrow \alpha''$ transformation is less favorable, and deformation is primarily governed by \{332\} or \{112\} twinning, again depending on loading conditions, consistent with TWIP-dominated behavior. At higher stability, as in Ti$_{4}$Nb at 300~K and Ti$_{2}$Nb at 77~K, \{332\} twinning is strongly suppressed, while \{112\} twinning and dislocation slip become dominant. Overall, these results demonstrate a clear correlation between deformation mode selection and $\beta$-phase stability. Importantly, our simulations show excellent agreement with experimental observations~\cite{zhao2021materials,lv2025novel,chen2018transitional,hanada1986effect,liang2020role,hanada1985deformation,yang2010evolution,zhan2016dynamic,shin2017phase,zhang2019plastic,gordin2020new}, as discussed in the next section.

\paragraph*{Experimental observations of deformation modes in Ti-Nb-based alloys}\label{para:Experimental}

\begin{figure}[htbp]
    \centering
    \makebox[\textwidth][c]{%
    \includegraphics[width=1\linewidth, clip, trim=0 0 0 0]{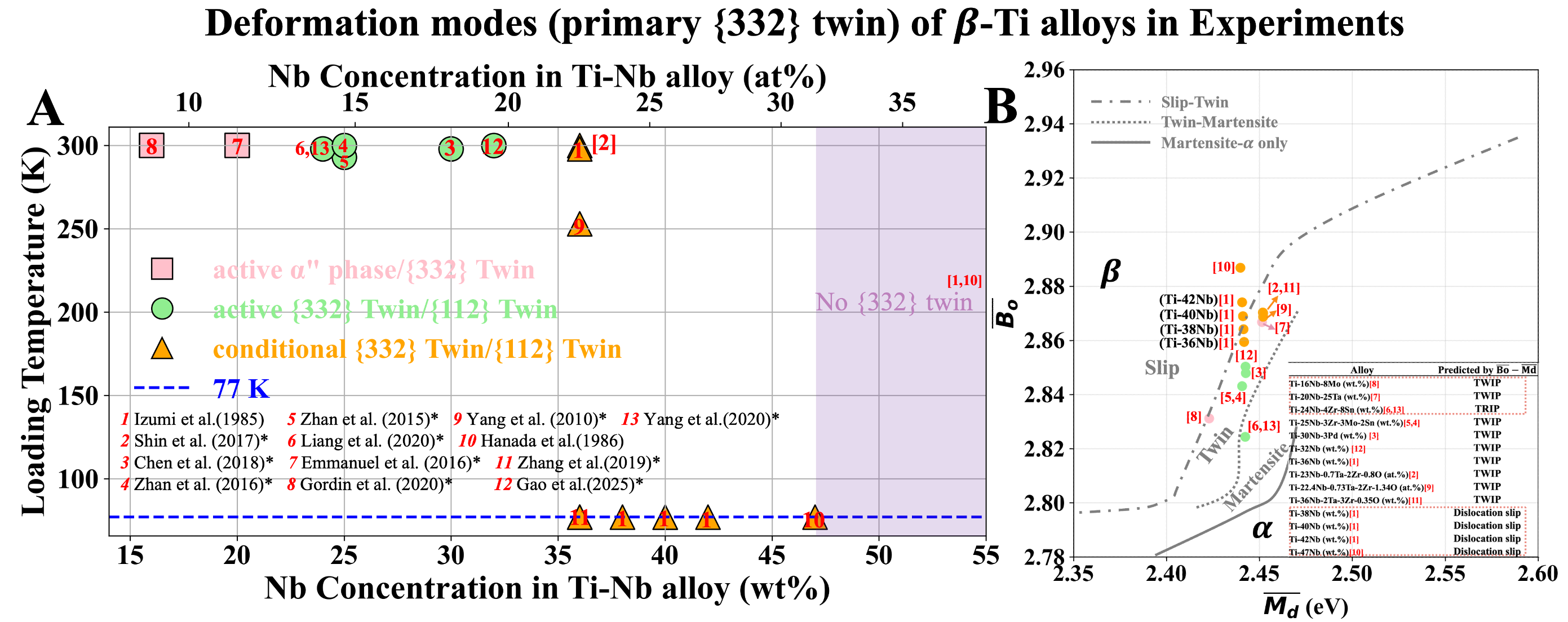}
    }
    \caption{\textbf{Temperature and composition conditions for \{332\} twin nucleation for Ti-Nb alloys in experiments ~\cite{hanada1986effect,hanada1985deformation,yang2010evolution,zhan2016dynamic,shin2017phase,chen2018transitional,liang2020role,zhang2019plastic,gordin2020new,yang2020plastic,gao2025manipulating,zhan2015deformation,bertrand2016deformation}.} In \textbf{A}, data points marked in pink rectangle (Nb concentration is less than 24 wt$\%$) indicates that dominant deformation mode is $\alpha"$ martensite and \{332\} twin. \{332\} twin (and small volume of \{112\} twin) can form insensitive to temperature and loading condition when Nb concentration is between 25 wt$\%$ and 36 wt$\%$ (data points in light green circle). \{332\} twin can form by carefully controlling loading condition under cryogenic temperatures (Nb concentration is between 36 wt$\%$ and 47 wt$\%$). Within this region, \{112\} twin and slip can also form and compete with \{332\} twin nucleation. When Nb concentration is larger than 47 wt$\%$, shown in purple region, \{332\} twin can not form in any condition and dislocation slip is the dominant deformation mode. As for alloys that marked with "*", Nb is the dominant $\beta$ stabilizer element in the composition but the alloy is not binary, they also include small amount of neutral elements and oxygen in addition. In \textbf{B}, we also present the predicted deformation modes of samples in \textbf{A} by $\overline{\text{Bo}}-\overline{\text{Md}}$ diagram. The definition of colors of data points in \textbf{B} are consistently based on the deformation modes observed in experiments in \textbf{A}.The experimental-fitted boundaries for different deformation modes are from Morinaga~\cite{morinaga2018molecular}. Clearly, compared to experimental observations of deformation modes in the samples shown in \textbf{A}, there are large deviation compared to $\overline{\text{Bo}}-\overline{\text{Md}}$ prediction, especially those marked in red rectangles.}
  \label{fig:Experimental}
\end{figure}

To assess consistency with experimental observations, we compile literature data on deformation modes in Ti–Nb alloys across a range of compositions and temperatures (Figure~\ref{fig:Experimental}A). Data points are color-coded by the reported dominant mechanism. In the low-stability regime (pink rectangle), experiments frequently show a mixture of $\beta \rightarrow \alpha''$ transformation and \{332\} twinning, consistent with our simulations of Ti$_5$Nb at 77~K that exhibit coexisting TRIP and TWIP (Figure~\ref{fig:Structural}A–F, Table~\ref{tab:deformationMode}). With slightly higher Nb content or temperature (green-circled region), \{332\} twinning predominates across 77–300~K~\cite{hanada1986effect}, though \{112\} twins occasionally appear under specific loading conditions, in agreement with our simulations for Ti$_5$Nb at 300~K and Ti$_4$Nb at 77~K. At higher $\beta$ stability (orange triangle region), \{332\} twins are limited to cryogenic conditions under carefully controlled loading, while deformation is otherwise dominated by dislocation slip~\cite{hanada1986effect}. Although experimental data here are sparse, our simulations of Ti$_4$Nb at 300~K and Ti$_2$Nb at 77~K show dominant \{112\} twinning and slip with suppressed \{332\} twinning. In the fully stable regime ($>$47 wt\% Nb, purple), no \{332\} twinning has been reported; deformation occurs almost exclusively by dislocation slip, as in conventional BCC metals. This trend matches our Ti$_2$Nb simulation at 300~K (Figure~\ref{fig:Structural}J–L, Table~\ref{tab:deformationMode}), where slip dominates and transformations or twinning are minimal. Overall, these comparisons demonstrate strong agreement between simulations and experiments, confirming the predictive fidelity of our framework.

Both our MD results (Table~\ref{tab:deformationMode}) and published experimental data (Figure~\ref{fig:Experimental}A) reveal a consistent trend in $\beta$-Ti alloys: as $\beta$-phase stability increases, the dominant deformation mechanisms shift from $\alpha''$ formation (TRIP) to \{332\} and \{112\} twinning (TWIP), and eventually to dislocation slip. This trend agrees with prior datasets~\cite{zhao2021materials,lv2025novel,coffigniez2024combination}. It can also be partially rationalized by the well-established $\overline{\text{Bo}}$–$\overline{\text{Md}}$ diagram~\cite{morinaga1988theoretical}, which uses concentration-averaged electronic structural parameters. Figure~\ref{fig:Experimental}B plots the corresponding $\overline{\text{Bo}}$ and $\overline{\text{Md}}$ values for alloys in Figure~\ref{fig:Experimental}A, with deformation boundaries based on Morinaga’s experimental fitting~\cite{morinaga2018molecular}. In this diagram, $\beta$ stability generally increases from lower right to upper left, and the observed TRIP $\rightarrow$ TWIP $\rightarrow$ slip sequence is broadly reproduced. The inset table in Figure~\ref{fig:Experimental}B summarizes the predicted modes. However, significant deviations remain, particularly in cases marked by dashed rectangles, where experimental results diverge from the $\overline{\text{Bo}}$–$\overline{\text{Md}}$ predictions. Notably, the diagram cannot distinguish between \{332\} and \{112\} twinning within the TWIP regime, nor does it capture the strong dependence of deformation behavior on temperature and loading conditions. These limitations highlight the need for more accurate, physically grounded models to describe deformation mode transitions under complex thermodynamic and mechanical environments.

These findings highlight a critical gap in our understanding of deformation in metastable $\beta$-Ti alloys. Specifically, three fundamental questions arise: (1) How does the free-energy landscape of diffusionless transformations at finite temperature shape deformation pathways, beyond the zero-Kelvin stability predictions of the $\overline{\text{Bo}}$–$\overline{\text{Md}}$ diagram? (2) What additional energetic parameters, beyond phase stability, govern the competition among deformation modes? (3) Can universal thermodynamic or kinetic metrics be identified to predict transitions between modes in the same alloy under different loading conditions? Addressing these questions requires moving beyond empirical descriptors toward an energetics-based framework. To this end, we analyze our MD simulations using advanced free-energy methods and crystallographic models linking twinning and diffusionless phase transformations. Our goal is to establish a predictive framework that connects temperature, composition, and loading conditions to deformation-mode selection—providing new principles for the design of $\beta$-Ti alloys with tailored mechanical performance through the controlled activation of competing mechanisms.

\paragraph*{$\beta$ phase stability and $\beta-\alpha"$ transformation reversibility}

We begin by exploring free-energy parameters related to $\beta$-phase stability to accurately capture their effects on deformation mode transition. Thus, we construct a free-energy landscape (FEL) for homogeneous $\beta \leftrightarrow \alpha''$ transformations using metadynamics simulations with PLUMED~\cite{plumed2019promoting,tribello2014plumed,bonomi2009plumed}. Following the Burgers path~\cite{burgers1934process}, we adopt two collective variables: the shuffle magnitude between two adjacent $(01\bar{1})$ planes along $[011]$ and the $b/a$ ratio, where $b$ and $a$ are the unit lengths of $\langle011\rangle$ and $\langle100\rangle$ in the $\beta$ phase. In the ideal $\beta$ phase (perfect BCC lattice), $b/a = \sqrt{2}$ and the shuffle magnitude is zero. In the orientation relationship between $\beta$ and the ideal HCP $\alpha$ phase, $b/a = \sqrt{3}$ and the shuffle magnitude is $b/6$, typically ranging from 0 to 0.1~nm depending on $\langle\bar{1}100\rangle$ length in HCP. The $\alpha''$ phase can thus be regarded as an intermediate state between the perfect BCC and HCP structures, with its equilibrium $b/a$ ratio and shuffle magnitude dependent on alloy composition and temperature. Further computational details are provided in the \textbf{Materials and Methods}.

\begin{figure}[htbp]
    \centering
    \makebox[\textwidth][c]{%
    \includegraphics[width=1\textwidth, clip=true, trim=0 0 0 0]{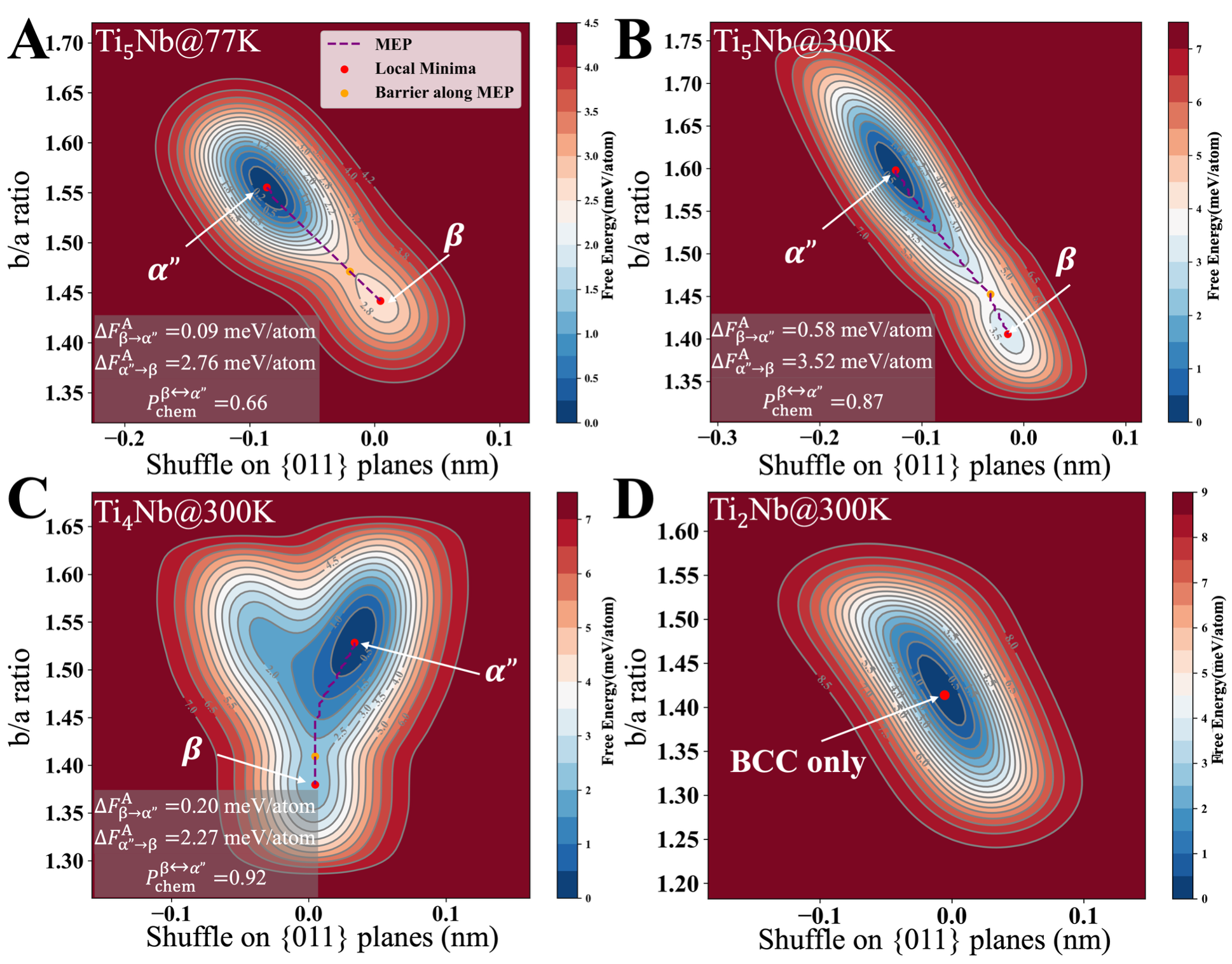}
    }
    \caption{\textbf{$\beta$-$\alpha"$ energy landscape for Ti alloys, including unstable BCC Ti-alloy ($\alpha$-Ti alloys), metastable BCC Ti-alloys and stable BCC Ti-alloy.} Here we choose two collective variables: shuffle magnitude between (011) planes along $[0\bar{1}1]$ directions and "b/a" ratios ("b" is the length of $\langle0\bar{1}1\rangle$ in BCC and  "a" is the length of $\langle100\rangle$ in BCC phase). Positions of BCC and $\alpha"$/$\alpha$ phase are marked in the figure. \textbf{A} is for Ti$_{15}$Nb at 300~K, an example of unstable BCC Ti-alloy ($\alpha$-Ti alloys). In this case, only one global minima located near $\alpha$ (HCP) phase and no local minima at BCC. \textbf{B-C} are cases of metastable BCC Ti-alloys: Ti$_{5}$Nb at 77~K and Ti$_{4}$Nb at 300~K. In this case, there are two minima corresponding to $\alpha"$ and BCC phase.\textbf{D} is the case of stable BCC Ti-alloys, which we can only observe one global minima at BCC phase and no local minima for $\alpha"$ phase. }
    \label{fig:ChemicalvsStrain}
\end{figure}

Figure~\ref{fig:ChemicalvsStrain}(A–D) shows the FELs of the $\beta \leftrightarrow \alpha''$ transformation in the unstable, metastable, and stable $\beta$-phase regions, respectively. By comparing alloy systems with different Nb concentrations and temperatures, we examine how $\beta$-phase stability influences $\beta \leftrightarrow \alpha''$ transformation behavior. To quantify the reversibility between the $\beta$ and $\alpha''$ phases, we introduce a parameter $P^{\beta \leftrightarrow \alpha''}_{\mathrm{chem}}$, defined as follows:

\begin{equation}
P^{\beta \leftrightarrow \alpha''}_{\mathrm{chem}} = \exp\left(-\frac{\left|\Delta F^{\textup{A}}_{\beta \leftrightarrow \alpha''}\right|}{K_{\text{B}}T}\right)
\label{eq:Boltzmann}
\end{equation}
Here 
\begin{equation}
\left| \Delta F^{\textup{A}}_{\beta \leftrightarrow \alpha''} \right| = \max\{  \Delta F^{\textup{A}}_{\beta \rightarrow \alpha''} , \Delta F^{\textup{A}}_{\alpha'' \rightarrow \beta} \}
\label{eq:Chem}
\end{equation}
Here, $\Delta F^{\textup{A}}_{\beta \rightarrow \alpha''}$ and $\Delta F^{\textup{A}}_{\alpha'' \rightarrow \beta}$ denote the activation free energies for the $\beta \rightarrow \alpha''$ and $\alpha'' \rightarrow \beta$ transformations along the minimum energy path (MEP), respectively. The parameter $P^{\beta \leftrightarrow \alpha''}_{\mathrm{chem}}$ ranges from 0 to 1, with lower values indicating reduced kinetic favorability for the reversible $\beta \leftrightarrow \alpha''$ transformation, and higher values indicating the opposite. The corresponding values are reported in the subfigures of Figure~\ref{fig:ChemicalvsStrain}. Details of the procedures for locating local minima and determining the MEP from the FEL are provided in the \textbf{Supplementary Materials}.

Figure~\ref{fig:ChemicalvsStrain}(A) shows the energy landscape of Ti$_5$Nb at 77~K, where the dominant deformation modes are $\alpha''$ formation or \{332\} twinning (Table~\ref{tab:deformationMode}). In this highly metastable state, the activation barrier for the $\beta \rightarrow \alpha''$ transformation is only 0.09~meV/atom, and $P^{\beta \leftrightarrow \alpha''}_{\mathrm{chem}}$ is 0.66—the lowest among all cases—indicating poor reversibility. Raising the temperature to 300~K (Figure~\ref{fig:ChemicalvsStrain}B) improves $\beta$ stability, increasing $P^{\beta \leftrightarrow \alpha''}_{\mathrm{chem}}$ to 0.87. Higher reversibility enables TWIP, allowing the system to escape the $\alpha''$ well under loading. The dominant modes here are \{112\} or \{332\} twinning, depending on loading conditions (Table~\ref{tab:deformationMode}). Composition also tunes the landscape: increasing Nb from Ti$_5$Nb to Ti$_4$Nb at 300~K (Figure~\ref{fig:ChemicalvsStrain}C) raises $P^{\beta \leftrightarrow \alpha''}_{\mathrm{chem}}$ to 0.92, with TWIP still dominant. At higher Nb (Ti$_2$Nb, 300~K; Figure~\ref{fig:ChemicalvsStrain}D), only a single global minimum corresponding to the $\beta$ phase remains. Here, the $\alpha''$ phase is unstable at zero stress, reversible transformation is negligible, and deformation occurs solely by dislocation slip (Table~\ref{tab:deformationMode}). 

In summary, these four representative cases demonstrate how $P^{\beta \leftrightarrow \alpha''}_{\mathrm{chem}}$ characterizes the effects of temperature and composition on $\beta \leftrightarrow \alpha''$ reversibility, correlating phase stability with the selection of TRIP, TWIP, or slip, consistent with Figure~\ref{fig:Experimental}. However, $P^{\beta \leftrightarrow \alpha''}_{\mathrm{chem}}$ alone cannot distinguish between twinning modes within the TWIP regime (e.g., Figures~\ref{fig:ChemicalvsStrain}B and C). Additional material properties beyond the FEL must therefore be considered, as discussed in the following section.

\paragraph*{Evaluations of misfit strain energies during reversible $\beta$-$\alpha"$ transformation}\label{para:TheoryHabitPlane}

We next consider factors beyond the FEL that determine the selection of specific twinning modes within the TWIP regime. Assuming twinning proceeds through a reversible $\beta \leftrightarrow \alpha''$ transformation, the preferred mode is that with the lowest nucleation barrier. This barrier includes four main contributions: the chemical driving force $\Delta G_{\text{chem}}^{\textup{A}}$, interfacial free energy $\Delta G_{\text{int}}^{\textup{A}}$, elastic strain energy from nucleus–matrix misfit $\Delta G_{\text{elast}}^{\textup{A}}$, and the external work $W_{\textup{ext}}$. The total nucleation barrier $\Delta G^{\textup{A}}$ can therefore be expressed as:

\begin{equation}
\Delta G^{\textup{A}} = \Delta G_{\text{chem}}^{\textup{A}} + \Delta G_{\text{int}}^{\textup{A}} + \Delta G_{\text{elast}}^{\textup{A}} -W_{\textup{ext}}
\label{eq:NucleationBarrier}
\end{equation}

Deformation twinning through reversible phase transformations typically forms coherent or semi-coherent twin boundaries~\cite{khachaturyan2013theory}. Consequently, $\Delta G_{\text{int}}^{\textup{A}}$ is usually negligible and not rate-controlling~\cite{khachaturyan2013theory}. Instead, the nucleation barrier is primarily governed by $\Delta G_{\text{chem}}^{\textup{A}}$, $\Delta G_{\text{elast}}^{\textup{A}}$ and external work $W_{\textup{ext}}$. The chemical term reflects the reversible $\beta \leftrightarrow \alpha''$ transformation barrier (related to $P^{\beta \leftrightarrow \alpha''}_{\mathrm{chem}}$ defined in the previous section), while the elastic penalty arises if the nucleus grows along a plane other than the invariant plane of the corresponding transformation~\cite{khachaturyan2013theory,nishiyama2012martensitic}. These invariant planes dictate the twin orientation relationship with the parent lattice and, consequently, the selection of twinning modes~\cite{khachaturyan2013theory,nishiyama2012martensitic}. Additionally, external work due to external loading can also alter the total twin nucleation barrier. Thus, the interplay between transformation reversibility and invariant-plane formation under loading condition is central to understanding deformation-mode selection in metastable $\beta$-Ti alloys.

\begin{figure}[htbp]
    \centering
    \makebox[\textwidth][c]{%
    \includegraphics[width=1\textwidth, clip=true, trim=0 0 0 0]{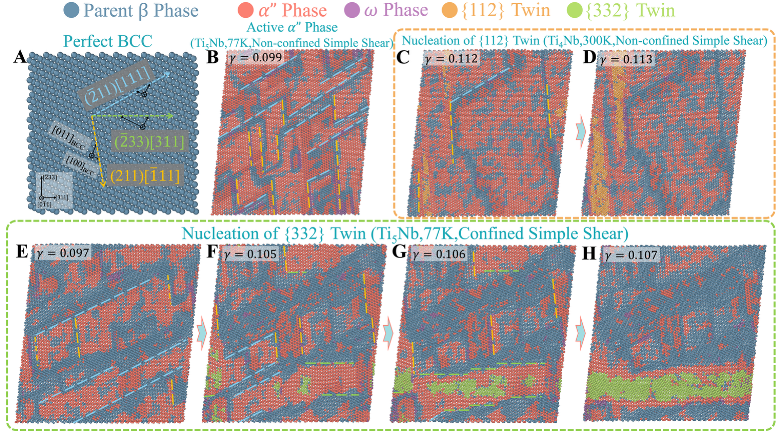}
    }
    \caption{\textbf{Facted interfaces of BCC and nucleated $\alpha"$ phase before nucleation of different deformation modes.} \textbf{(A)} is illustration on the all conventional coherent $\text{$\beta$}$$\parallel$$\alpha"$ interfaces (marked in orange and blue) consistent with the BOR and defined by \{211\} planes with in-plane $\langle\bar{1}11\rangle$ directions, lying in the [311]–[$\bar{2}$33] zone. We also mark the ($\bar{2}$33) plane along [311] direction (green line) that closely align to the \{332\} twin boundary formed under shear loading. The coordinate system in \textbf{(A)} is consistent with that of initial state of simulation box in \textbf{(B-H)}. \textbf{(B)} is the snapshot when the system close to the stage of $\alpha"$ phase nucleation saturation. Corresponding setting is consistent with the case in Figure ~\ref{fig:Structural} (A-D). \textbf{(C-D)} are snapshots of critical stage of \{112\} twin nucleation in the case of Figure ~\ref{fig:Structural} (I-L). \textbf{(E-H)} are the snapshots of \{332\} twin nucleation process in the case of Figure ~\ref{fig:Structural} (E-H). Here we want to present how the interfaces of $\text{$\beta$}$$\parallel$$\alpha"$  (bottom of the figures) tilt from blue lines to green lines that closely align with \{332\} twinning plane. Definitions on the colors of atoms are consistent with the definition in Figure ~\ref{fig:Structural}. Lines in \textbf{(B-H)} mark some facted interfaces between BCC and nucleated $\alpha"$ phase. Lines with the same color as \textbf{A)} correspond to closely related orientation.}
    \label{fig:FacetPlanes}
\end{figure}

Within this framework, twinning via reversible transformations proceeds in two steps: (1) $\beta$ (matrix) $\rightarrow \alpha''$, and (2) $\alpha'' \rightarrow \beta$ (twin lattice). Each step ideally occurs along a distinct invariant plane, the habit plane for the phase transformation and the twin plane for twinning, thereby minimizing misfit strain energy~\cite{zhang2016faceted}. Along the lowest free-energy pathway, $\alpha''$ nucleates and grows along the habit plane in step (1). After saturation, it transforms back to $\beta$ in step (2), initially along the same habit plane and ultimately along the twin plane between the $\beta$ matrix and the $\beta$ twin lattice. Under strong local driving forces (e.g., stress concentrations near nucleation sites), however, the $\alpha''$ phase in step (1) may deviate from its ideal invariant-plane orientation to accelerate kinetics, incurring higher misfit energy that can significantly contribute to $\Delta G^{\textup{A}}$. For example, experimental characterizations of $\beta \rightarrow \alpha \rightarrow \alpha''$ transformations often reveal semi-coherent interfaces that follow the Burgers Orientation Relationship (BOR): $\{1\bar{1}2\}_{\beta} \parallel \{0\bar{1}10\}_{\text{HCP}} \parallel \{110\}_{\alpha''}$ and $\langle\bar{1}11\rangle_{\beta} \parallel \langle\bar{2}110\rangle_{\text{HCP}} \parallel \langle\bar{1}10\rangle_{\alpha''}$~\cite{wang2023coherent}. To evaluate the resulting misfit strain energies, we next analyze atomistic structures, with particular focus on the $\beta \parallel \alpha''$ interfaces at the twin nucleation stage shown in Figure~\ref{fig:FacetPlanes}.
 
Figure~\ref{fig:FacetPlanes}(A) illustrates the $(\bar{2}33)_{\beta}$ plane (green), where shear deformation is later applied along the $[311]_{\beta}$ direction; and two typical $\beta \parallel \alpha''$ semi-coherent interfaces (blue and yellow) lying in the $[\bar{2}33]_{\beta}$–$[311]_{\beta}$ plane that follow the BOR introduced earlier. Figures~\ref{fig:FacetPlanes}(B–H) show representative faceted interfaces observed at different stages: near $\alpha''$ saturation (Figure~\ref{fig:FacetPlanes}(B), corresponding to Ti$_5$Nb at 77~K under NSS loading shown in Figure~\ref{fig:Structural}B), during \{112\} twin nucleation (Figures~\ref{fig:FacetPlanes}(C–D), corresponding to Ti$_4$Nb at 300~K under NSS loading shown in  Figure~\ref{fig:Structural}H), and during \{332\} twin nucleation (Figures~\ref{fig:FacetPlanes}(E–H), corresponding to Ti$_5$Nb at 77~K under CSS loading plotted in Figure~\ref{fig:Structural}E). In the case of $\alpha''$ formation plotted in Figure~\ref{fig:FacetPlanes}(B), $\alpha''$ nucleation and growth is strongly favored for Ti$_5$Nb at 77~K under NSS loading. The large $\alpha''$ volume fraction promotes the formation of multiple coherent and semi-coherent interfaces to minimize interfacial energy. Two of the most prominent align closely with \{211\} planes [Figure~\ref{fig:FacetPlanes}(A)], consistent with the BOR as well as prior experimental and computational studies~\cite{murzinova2021effect,shi2012predicting,wang2023coherent}. With further shear, only small \{112\} twin nuclei are observed, reflecting both the low reversibility and the tendency of this loading condition to stabilize $\alpha''$. For \{112\} twin nucleation (Figures~\ref{fig:FacetPlanes}(C–D)), a long, flat $\beta \parallel \alpha''$ interface forms nearly parallel to the $(211)_{\beta}$ plane along $[\bar{1}11]_{\beta}$ (marked in orange), corresponding to the theoretical invariant plane of the \{112\} twin shown in Figure~\ref{fig:Structural}(J).

\begin{figure}[htbp]
    \centering
    \makebox[\textwidth][c]{%
    \includegraphics[width=0.6\textwidth, clip=true, trim=0 0 0 0]{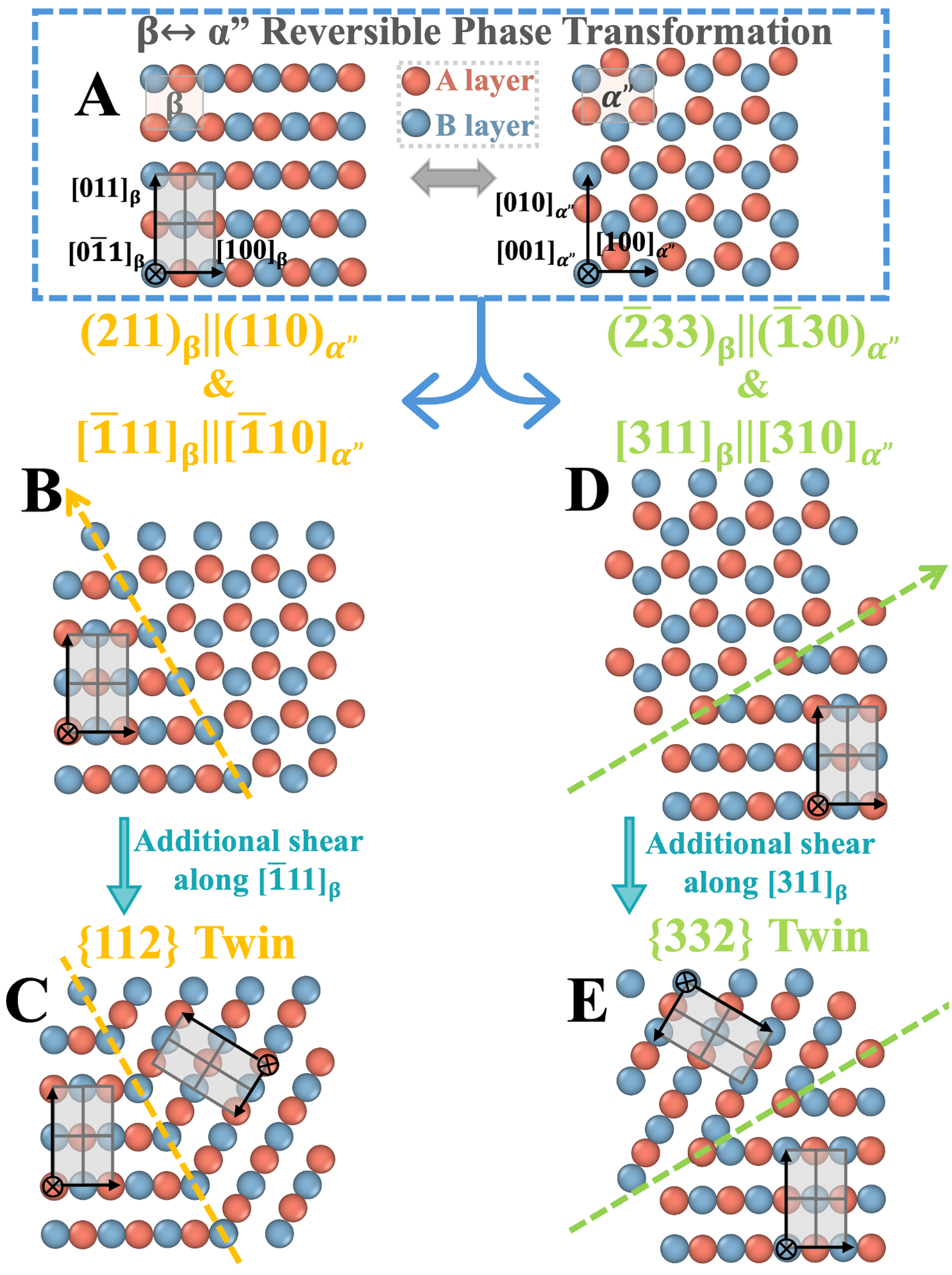}
    }
    \caption{\textbf{Schematic draw on BCC-to-$\alpha"$ phase transformation and two BCC-$\alpha"$ interphase boundary align with twinning plane.} \textbf{A} illustrate BCC-$\alpha"$ phase transformation pathway by shuffling (0$\bar{1}$1) planes along [011] direction. Color of atoms indicating ABAB stacking. In addition to shuffle, the lattice also undergo lateral strain along [100] directions during transformation. To simplify illustration, we do not draw here. $\alpha"$ phase is similar to HCP phase, and the real atomic structure of $\alpha"$ phase varies with different metastable alloys under different temperature, we draw perfect HCP here as an example. \textbf{B} and \textbf{D} are two theoretical calculated BCC-$\alpha"$ interphase boundaries align with twinning planes during BCC-$\alpha"$ phase transformation. Orange and green arrows indicate the directions that parallel to the interface. \textbf{C} and \textbf{E} illustrate two types of BCC twin boundary structure after $\alpha"$ phase transfering back to BCC phases under shear along the directions marked in blue arrows. \textbf{A-B-C} is the pathway for \{112\} twin formation by BCC-to-$\alpha"$ phase reversible phase transformation and \textbf{A-D-E} is the pathway for \{332\} twin formation by BCC-to-$\alpha"$ phase reversible phase transformation. In real cases (both in MD and experiments), there are phase transformation behaviors along twin boundaries. Since we here is to illustrate structural relationships of $\alpha"$, twin and matrix, we do not draw the real twin boundary structure. The illustrations and snapshots of twin boundaries are in \textbf{Supplementary Materials}.}
    \label{fig:schematic}
\end{figure}

In the case of \{332\} twin nucleation [Figure~\ref{fig:FacetPlanes}(E–H)], we observe uncommon faceted $\beta \parallel \alpha''$ interfaces (marked in green) oriented near the $(\bar{2}33)_{\beta}$ plane, which later becomes the invariant plane of the nucleated \{332\} twin. Initially, however, the $(\bar{2}33)_{\beta}$ plane along $[311]_{\beta}$ is not the preferred $\beta \parallel \alpha''$ interface [Figure~\ref{fig:FacetPlanes}(E)]. When local \{332\} twin domains form within the $\alpha''$ band at the bottom of the cell, the band adopts a curved shape [Figure~\ref{fig:FacetPlanes}(F–G)], and the emerging \{332\} twins tend to reorient the $\beta \parallel \alpha''$ interface toward alignment with the \{332\} twinning plane (green line), progressively flattening the interface. As additional \{332\} twin domains form, the faceted $\beta \parallel \alpha''$ interface gradually becomes parallel to the \{332\} twinning planes [Figure~\ref{fig:FacetPlanes}(H)]. The dynamic reorientation of the interface from $(\bar{2}11)_{\beta}$ to $(\bar{2}33)_{\beta}$ is captured in the \textbf{Supporting Movie}. To our knowledge, this represents the first atomic-scale observation of dynamic \{332\} twin nucleation from the $\alpha''$ phase, consistent with in-situ experiments~\cite{lai2016mechanism}. This finding also confirms our assumption that the formation of the twin invariant plane requires the $\beta \parallel \alpha''$ interface to align with the twin plane. Considering the two critical steps of twin nucleation discussed above, we regard the alignment of $\alpha''$ along faceted $\beta \parallel \alpha''$ interfaces toward the twin-invariant orientation as an intermediate step connecting them. Accordingly, in our model, the in-plane misfit strain energy density of coherent or semi-coherent $\beta \parallel \alpha''$ interfaces aligned with twin-invariant planes is treated as the dominant contribution to the misfit strain energy governing twin nucleation and, ultimately, twin selection.
As shown in Figure~\ref{fig:schematic}, crystallographic analysis demonstrates that the $\beta \parallel \alpha''$ interfaces and the associated twin-invariant plane orientations observed in Figure~\ref{fig:FacetPlanes} can be obtained through the reversible $\beta \leftrightarrow \alpha''$ phase transformation along the well-defined Burgers path for the transformation between $\beta$ and $\alpha$ phase in the HCP structure, because the $\alpha''$ phase can be treated as a distorted $\alpha$ phase deviated from the ideal $c/a$ ratio and the perfect hexagonal basal plane ~\cite{burgers1934process,gao2020intrinsic}. In Figure~\ref{fig:schematic}(A), different atom colors highlight the ABAB stacking sequence along $[0\bar{1}1]_{\beta}$, while the $\beta \rightarrow \alpha''$ transformation involves a homogeneous compressive strain along $[100]_{\beta}$ together with an additional atomic shuffle of $(0\bar{1}1)_{\beta}$ planes along $[011]_{\beta}$. Based on the observations in Figure~\ref{fig:FacetPlanes}, once the $\alpha''$ phase nucleates and grows into a large band or grain within the parent $\beta$ matrix, two key $\beta \parallel \alpha''$ coherent/semi-coherent interfaces aligned with different twin-invariant planes can be identified (Figures~\ref{fig:schematic}(B–E)). Figures~\ref{fig:schematic}(B–C) correspond to the formation of the \{112\} twin, consistent with the MD results in Figures~\ref{fig:FacetPlanes}(C–D), while Figures~\ref{fig:schematic}(D–E) correspond to the formation of the \{332\} twin, consistent with Figures~\ref{fig:FacetPlanes}(E–H). From a crystallographic perspective, and following the Hadamard jump condition~\cite{christian1995deformation,gao2020intrinsic}, the intermediate $\alpha''$ phase connecting the parent $\beta$ variant and the twin variant (\{112\} or \{332\}) exhibits the following orientation relationship (OR) with the $\beta$ phase:
\begin{equation}
\begin{aligned}
\text{For }\{112\}\text{ twin: } (211)_{\mathrm{\beta}} &\parallel (110)_{\alpha''}, \quad [\bar{1}11]_{\mathrm{\beta}} \parallel [\bar{1}10]_{\alpha''}, \\
\text{For }\{332\}\text{ twin: } (\bar{2}33)_{\mathrm{\beta}} &\parallel (\bar{1}30)_{\alpha''}, \quad [311]_{\mathrm{\beta}} \parallel [310]_{\alpha''}.
\end{aligned}
\label{eq:OR_variant}
\end{equation}


Based on the above orientation relationships, we next estimate the in-plane misfit strain energy along the $\beta \parallel \alpha''$ interfaces, assuming that the local interfaces behave orthotropically. Thus, the in-plane stress–strain relation (neglecting out-of-plane components) follows the orthotropic form~\cite{jones2018mechanics}. Furthermore, using Eshelby’s theory~\cite{eshelby1959elastic} in combination with the orthotropic stress–strain relation, we derive the general expression for the in-plane misfit strain energy density, denoted as $W^{\text{misfit}}_{\text{twin}}$ (in units of eV/atom), as:
\begin{equation}
\begin{aligned}
W^{\text{misfit}}_{\text{twin}} = \tfrac{1}{2} 
\left( 
\frac{E_{1}}{1 - \nu_{12}\nu_{21}}\varepsilon_{11}^2 
+ \frac{E_{2}}{1 - \nu_{12}\nu_{21}}\varepsilon_{22}^2 
+ 2\frac{\nu_{12}E_{2}}{1 - \nu_{12}\nu_{21}}\varepsilon_{11}\varepsilon_{22} 
+ G_{12}\gamma_{12}^2
\right)\Omega_{\beta}
\end{aligned}
\label{eq:MisfitGeneral}
\end{equation}
where direction “1” denotes the longitudinal in-plane axis ($[\bar{1}11]_{\beta}$ or $[311]_{\beta}$ in Equation~\ref{eq:OR_variant}), and direction “2” denotes the transverse in-plane axis (given by the cross product of the plane normal and the longitudinal direction in Equation~\ref{eq:OR_variant}); in this study, the transverse in-plane axis is $[0\bar{1}1]_{\beta}$, which is parallel to the Z axis. $E_{1}$ and $E_{2}$ represent the Young’s moduli along the longitudinal and transverse directions, respectively; $\nu_{12}$ and $\nu_{21}$ are the corresponding Poisson’s ratios, with $\nu_{12}$ describing the transverse strain induced by longitudinal stress and $\nu_{21}$ the reverse. $G_{12}$ is the in-plane shear modulus, $\gamma_{12}$ the in-plane shear strain, $\varepsilon_{11}$ and $\varepsilon_{22}$ the misfit strains along the longitudinal and transverse directions, and $\Omega_{\beta}$ the atomic volume of the $\beta$ phase. The in-plane elastic constants, Poisson’s ratios, and misfit strains for the $\beta \parallel \alpha''$ interfaces associated with the \{112\} and \{332\} twins are summarized in the \textbf{Supplementary Materials}. The misfit strain energy is defined relative to the $\beta$ phase, since the nucleated $\alpha''$ domains act as inclusions embedded within the parent $\beta$ matrix, and the surrounding $\beta$ lattice accommodates the misfit. Accordingly, all elastic constants and Poisson’s ratios used in this study correspond to the $\beta$ parent phase. The specific $\beta \parallel \alpha''$ interfaces used to calculate $W^{\text{misfit}}_{\text{twin}}$ for both twin modes, $W^{\text{misfit}}_{\text{\{332\} twin}}$ and $W^{\text{misfit}}_{\text{\{112\} twin}}$, follow the orientation relationships defined in Equation~\ref{eq:OR_variant}.

\paragraph*{Twin nucleation metric}\label{para:Chemical_Strain_Correlation} 

Fig.~\ref{fig:FacetPlanes} and Fig.~\ref {fig:schematic} suggest approximate methods to describe the twin nucleation path through the $\beta \leftrightarrow \alpha''$ reversible phase transformation for both \{112\} and \{332\} twins. Based on these descriptions of reaction paths and each term in Eq.\ref{eq:NucleationBarrier} for the nucleation barrier, we here now propose a twin nucleation metric (TNM):
\begin{equation}
\begin{aligned}
\Delta g^{*}_{\text{twin}}=\left| \Delta F^{\textup{A}}_{\beta \leftrightarrow \alpha''} \right|+W^{\text{misfit}}_{\text{twin}}
\end{aligned}
\label{eq:EffNuc}
\end{equation}
Here $\Delta g^{*}_{\text{twin}}$ characterizes the total free energy barrier per atom if the twinning is achieved by the reversible transformation between BCC and $\alpha"$ phase with the critical intermediate $\alpha"$ phase growing along the twin invariant planes. $\left| \Delta F^{\textup{A}}_{\beta \leftrightarrow \alpha''} \right|$ defined in Eq.~\ref{eq:Chem} (correlate $\Delta G_{\text{chem}}^{\textup{A}}$ in Eq.~\ref{eq:NucleationBarrier}) characterizes the chemical free energy contribution to the barrier and is mainly dependent on temperature and composition. $W^{\text{misfit}}_{\text{twin}}$ from Eq.~\ref{eq:MisfitGeneral} estimates the misfit strain energy costs (correlate $\Delta G_{\text{elas}}^{\textup{A}}$ and $W_{\textup{ext}}$ in Eq.~\ref{eq:NucleationBarrier}), which are dependent on temperature, composition, and loading conditions. Then we use it to evaluate the twin nucleation behaviors we summarize in Table~\ref{tab:deformationMode} and to investigate how and why temperature, composition, and loading condition tune the twinnability.

\textbf{Table}~\ref{tab:StrainCalc} summarizes the critical energetic descriptors for twinning nucleation behavior, including those defined in Eq.~\ref{eq:EffNuc}, for four Ti–Nb alloy systems prior to loading (the corresponding dominant twinning modes after loading are listed in the rightmost column): Ti$_{5}$Nb at 77~K, Ti$_{5}$Nb at 300~K, Ti$_{4}$Nb at 77~K, and Ti$_{4}$Nb at 300~K. In addition, one experimental dataset (Ti–10Mo~wt.\% at 300~K) beyond the Ti–Nb system is also included. Detailed procedures for calculating the equilibrium lattice constants of the BCC and $\alpha''$ phases are described in the \textbf{Materials and Methods}. $P^{\beta \leftrightarrow \alpha''}_{\mathrm{chem}}$, $\Delta g^{*}_{\text{\{112\} twin}}$, and $\Delta g^{*}_{\text{\{332\} twin}}$ are obtained from $\left| \Delta F^{\textup{A}}_{\beta \leftrightarrow \alpha''} \right|$, $w^{\mathrm{misfit}}_{\mathrm{\{112\} twin}}$, and $w^{\mathrm{misfit}}_{\mathrm{\{332\} twin}}$ according to Equation~\ref{eq:OR_variant} and Equation~\ref{eq:MisfitGeneral}. The magnitudes of these energetic parameters fall within a reasonable range compared with experimental results for Ti alloys~\cite{zhao2021materials}. In all cases shown in \textbf{Table}~\ref{tab:StrainCalc}, the misfit strain energy density is comparable to, or even higher than, the activation energy barrier for the phase transformation. This finding indicates that, in some cases, the elastic strain energy can be more critical than the chemical free energy (i.e., BCC phase stability) in determining the twinnability and the selection of the preferred twinning mode.

In the case of Ti$_{5}$Nb at 77~K, $P^{\beta \leftrightarrow \alpha''}_{\mathrm{chem}}$ is relatively low compared to the other cases, indicating that the reversibility between the BCC and $\alpha''$ phases is limited. As confirmed by the MD simulations, the $\alpha''$ phase can be retained even under severe deformation in specific loading conditions. For Ti$_{5}$Nb at 300~K and Ti$_{4}$Nb at 77~K, $P^{\beta \leftrightarrow \alpha''}_{\mathrm{chem}}$ is closer to 1, suggesting a higher reversibility between the BCC and $\alpha''$ phases than in Ti$_{5}$Nb at 77~K. Consequently, the $\alpha''$ phase acts as an intermediate stage, and twinning becomes easier to initiate under applied loading. In both cases, the twin nucleation metric (TNM) $\Delta g^{*}_{\text{\{112\} twin}}$ is about 2.5–3.5~meV/atom lower than $\Delta g^{*}_{\text{\{332\} twin}}$, indicating that the \{112\} twin is intrinsically more favorable. However, external loading conditions can compensate for the difference in $\Delta g^{*}_{\text{twin}}$, making \{332\} twin nucleation energetically competitive. Therefore, both \{112\} and \{332\} twins may form depending on the specific loading environment. For Ti$_{4}$Nb at 300~K, the high $P^{\beta \leftrightarrow \alpha''}_{\mathrm{chem}}$ (0.92) ensures a strong propensity for twin nucleation. In this case, $\Delta g^{*}_{\text{\{112\} twin}}$ is significantly smaller (by approximately 5~meV/atom) than $\Delta g^{*}_{\text{\{332\} twin}}$, indicating that even under loading conditions that nominally favor \{332\} twinning, only \{112\} twins are likely to form. Conversely, for Ti–10Mo~wt.\% at 300~K, $\Delta g^{*}_{\text{\{332\} twin}}$ is about 8.6~meV/atom lower than $\Delta g^{*}_{\text{\{112\} twin}}$, which explains why only \{332\} twins are observed experimentally, despite the complexity of the loading conditions.  In summary, $\Delta g^{*}_{\text{twin}}$ in the undeformed state, when combined with $P^{\beta \leftrightarrow \alpha''}_{\mathrm{chem}}$, reflects the intrinsic tendency for twinning and twin mode selection—independent of external loading effects—to some extent. We next investigate how loading conditions influence deformation mode selection by analyzing the evolution of $\Delta g^{*}_{\text{twin}}$ during dynamic shear loading.

\begin{figure}[htbp]
    \centering
    \makebox[\textwidth][c]{%
    \includegraphics[width=1\textwidth, clip=true, trim=0 0 0 0]{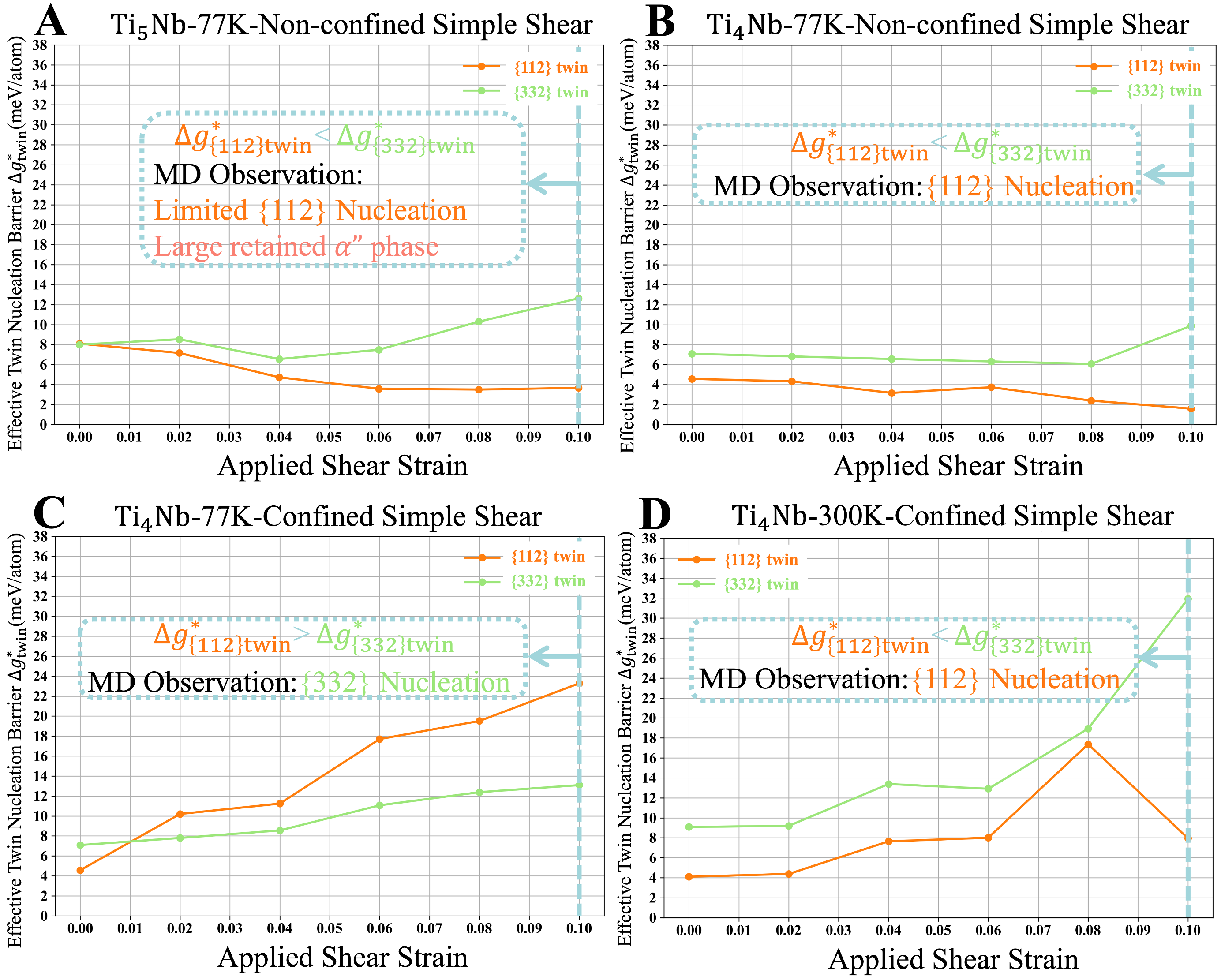}
    }
    \caption{\textbf{Dynamic change of twin nucleation metric, $\Delta g^{*}_{\text{twin}}$ under different stress/strain conditions for different alloy systems.} In all cases, the volume of nucleated $\alpha"$ phase saturates when applied shear strain is around 0.1. And twins starts to nucleate right after. The orange curves represent $\Delta g^{*}_{\text{\{112\}twin}}$ under loading conditions while the green curves correspond to $\Delta g^{*}_{\text{\{332\}twin}}$ under loading conditions. The alloy composition, temperature and loading type are at top of each subfigure. We divide four cases: \textbf{A-B}, \textbf{C-D}, \textbf{E-F} and \textbf{G-H}. Within these four sets, the temperature and composition are the same but under different loading conditions. In each subfigure, we compare the magnitude of $\Delta g^{*}_{\text{\{112\}twin}}$ and $\Delta g^{*}_{\text{\{332\}twin}}$ at critical stage (external shear strain is 0.1,marked in light blue lines). Meanwhile, we mark the MD observations on the type of twin about to nucleate. All the snapshots of final deformed simulation box are in \textbf{Supplementary Materials}.}
    \label{fig:LoadStrain}
\end{figure}

Figure~\ref{fig:LoadStrain} shows the evolution of $\Delta g^{*}_{\text{twin}}$ during dynamic loading for different alloy systems. Details of the $\Delta g^{*}_{\text{twin}}$ calculation under deformed states are provided in the \textbf{Materials and Methods}, and analyses of atomic structure evolution and deformation mode nucleation for the cases shown in Figure~\ref{fig:LoadStrain} are included in the \textbf{Supplementary Materials}. During shear deformation, $\beta$–$\alpha''$ phase transformations are observed in all cases, and the volume fraction of the nucleated $\alpha''$ phase saturates when the applied shear strain reaches approximately 0.1. Subsequently, the nucleated $\alpha''$ phase transforms into different deformation modes, depending on the temperature, composition, and corresponding stress/strain conditions. The detailed simulation parameters are labeled at the top of each subplot.  

For Ti$_{5}$Nb at 77~K under non-confined simple shear (Figure~\ref{fig:LoadStrain}(A)), $\Delta g^{*}_{\text{\{112\} twin}}$ and $\Delta g^{*}_{\text{\{332\} twin}}$ are initially similar. Upon applying shear strain, the loading condition alters these values differently: $\Delta g^{*}_{\text{\{112\} twin}}$ decreases, while $\Delta g^{*}_{\text{\{332\} twin}}$ increases. At the critical stage, where the $\alpha''$ phase fraction reaches saturation, $\Delta g^{*}_{\text{\{112\} twin}}$ becomes approximately 8~meV/atom lower than $\Delta g^{*}_{\text{\{332\} twin}}$, making \{112\} twin nucleation more favorable. However, as discussed in Table~\ref{tab:StrainCalc}, the reversibility between the BCC and $\alpha''$ phases is low. Consistently, the MD simulations show that the dominant deformation mode remains the $\alpha''$ phase, accompanied by a few localized \{112\} twin nuclei. Even when \{112\} twins form, their volume fraction remains small, and the system largely retains the $\alpha''$ structure. The limited twin volume fraction can also be attributed to the large barrier associated with twin growth~\cite{chen2024stability}, which lies beyond the scope of this study but will be addressed in future work.

If we increase the Nb concentration from Ti$_{5}$Nb to Ti$_{4}$Nb while keeping the same temperature and loading condition [Figure~\ref{fig:LoadStrain}(B)], $P^{\beta \leftrightarrow \alpha''}_{\mathrm{chem}}$ increases (as already shown in Table~\ref{tab:StrainCalc}), indicating higher reversibility between the BCC and $\alpha''$ phases. Meanwhile, the overall trend in the evolution of $\Delta g^{*}_{\text{twin}}$ remains similar to that in Figure~\ref{fig:LoadStrain}(A). In this case, the $\alpha''$ phase is not retained, and evident \{112\} twin nucleation behavior is observed. In contrast, if we keep the same composition and temperature but change the loading condition (comparing Figure~\ref{fig:LoadStrain}(B) and (C)), the twinning nucleation behavior changes dramatically. The loading condition in Figure~\ref{fig:LoadStrain}(C) increases both $\Delta g^{*}_{\text{\{112\} twin}}$ and $\Delta g^{*}_{\text{\{332\} twin}}$, but the increase in $\Delta g^{*}_{\text{\{112\} twin}}$ is much larger. As a result, although $\Delta g^{*}_{\text{\{112\} twin}}$ is initially about 2.5~meV/atom lower than $\Delta g^{*}_{\text{\{332\} twin}}$, under loading it becomes roughly 10~meV/atom higher at the stage where the $\alpha''$ phase fraction saturates. Consequently, \{332\} twin nucleation becomes much more favorable than \{112\} twinning, consistent with the MD observations.  

Temperature also affects twinning mode selection, as revealed by the evolution of $\Delta g^{*}_{\text{twin}}$. Comparing the cases in Figure~\ref{fig:LoadStrain}(C) and (D), where alloy composition and loading conditions are identical but temperature differs, we find that in Figure~\ref{fig:LoadStrain}(D), $\Delta g^{*}_{\text{\{112\} twin}}$ is initially much smaller than $\Delta g^{*}_{\text{\{332\} twin}}$ compared to the case in Figure~\ref{fig:LoadStrain}(C). Even though further loading significantly increases $\Delta g^{*}_{\text{\{112\} twin}}$, $\Delta g^{*}_{\text{\{332\} twin}}$ remains higher at the critical stage. As a result, \{112\} twinning, rather than \{332\}, is still observed.  In summary, consistent with the discussion in Table~\ref{tab:StrainCalc}, a large nucleation energy difference between $\Delta g^{*}_{\text{\{112\} twin}}$ and $\Delta g^{*}_{\text{\{332\} twin}}$ under stress-free conditions limits the influence of loading effects, and the twinning mode preferentially follows the pathway with the lower $\Delta g^{*}_{\text{twin}}$.

\section*{Discussion}
\label{discuss}

\begin{figure}[htbp]
\centering
    \makebox[\textwidth][c]{%
    \includegraphics[width=1\textwidth, clip=true, trim=0 0 0 0]{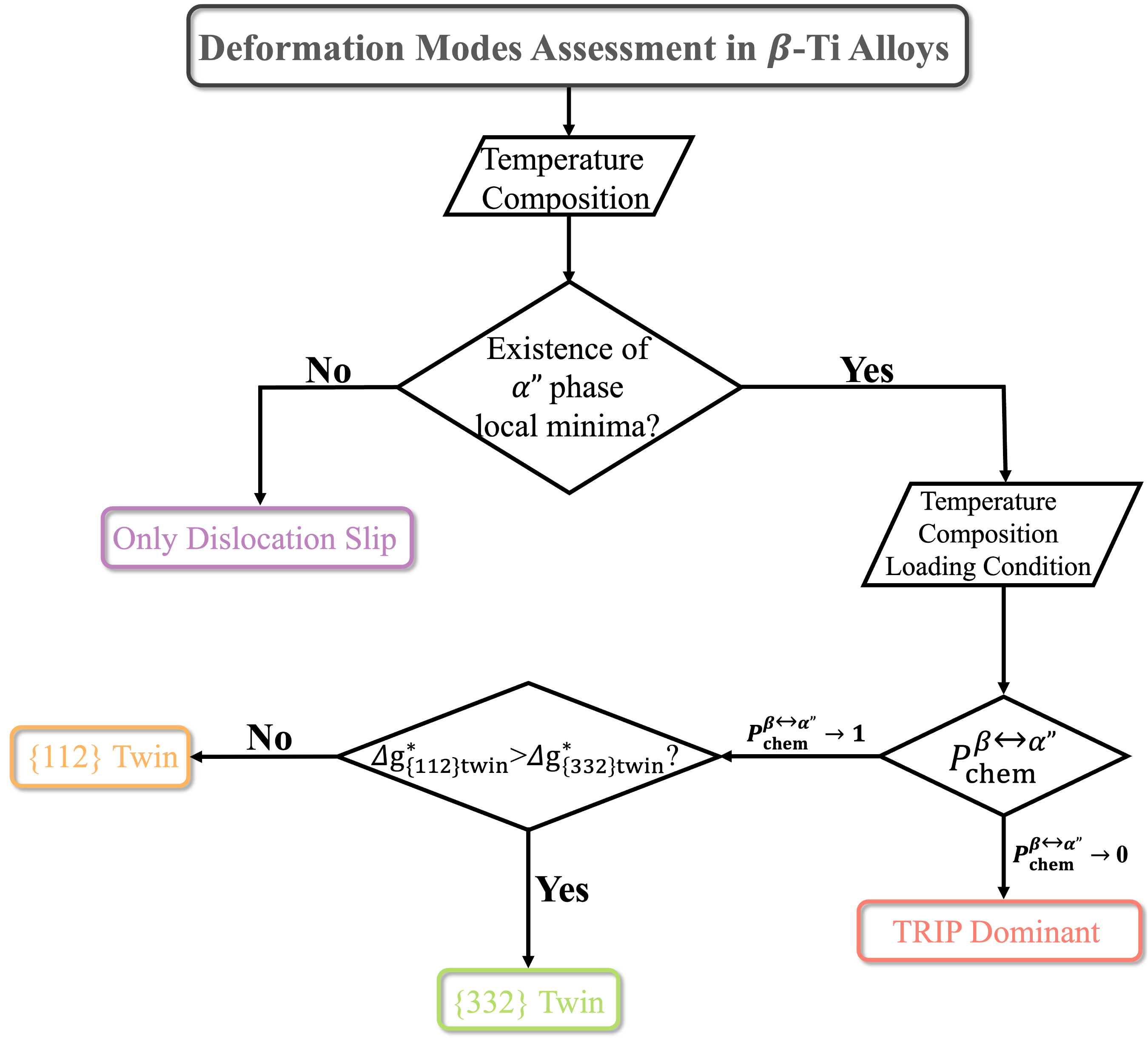}
    }
\caption{A conceptual flowchart for criteria of deformation modes selection in metastable $\beta$-Ti alloys.}
\label{fig:flowchart}
\end{figure}

So far, our proposed $\Delta g^{*}_{\text{twin}}$ illustrates how temperature, composition, and loading conditions influence twinning mode selection in the metastable BCC region, while $P^{\beta \leftrightarrow \alpha''}_{\mathrm{chem}}$ indicates how temperature and alloy composition affect the competition among TWIP, TRIP, and dislocation slip. By integrating these two parameters, we introduce a conceptual flowchart in Fig.~\ref{fig:flowchart} to illustrate how deformation modes can be tuned via temperature, composition, and loading conditions. We first examine whether the $\alpha''$ phase exists at the stress-free state to determine when dislocation slip dominates, represented by the left branch of the flowchart. In this study, Ti$_{2}$Nb at 300~K serves as a representative example of this regime. If the $\alpha''$ phase can exist at the stress-free state and corresponds to a local minimum in the FEL, we then evaluate the reversibility between the BCC and $\alpha''$ phases and the twin nucleation metric to identify the competition among TRIP, \{112\} twinning, and \{332\} twinning. This corresponds to the right branch of the flowchart. We first assess $P^{\beta \leftrightarrow \alpha''}_{\mathrm{chem}}$. Although no strict threshold is defined here, a low reversibility (with $P^{\beta \leftrightarrow \alpha''}_{\mathrm{chem}}$ close to 0) indicates that twinning is difficult to nucleate via reversible phase transformations, and the $\alpha''$ phase can be retained under certain loading conditions. Conversely, when the reversibility between the BCC and $\alpha''$ phases is high, twins can form more readily. Under this condition, we then compare $\Delta g^{*}_{\text{twin}}$ for \{112\} and \{332\} twins. Although loading conditions can strongly modify the twin nucleation barriers, particularly near the critical stage of $\alpha''$ phase saturation, a large intrinsic (i.e., under the loading-free condition) difference in $\Delta g^{*}_{\text{twin}}$ between the two twinning modes makes the mode with the lower $\Delta g^{*}_{\text{twin}}$ value more favorable and less sensitive to external loading. Representative examples include the \{112\} twin in Ti$_{4}$Nb at 300~K and the \{332\} twin in Ti–10Mo~wt.\% at 300~K.  

When $\Delta g^{*}_{\text{twin}}$ for the \{112\} and \{332\} twins are comparable, the determining factor is no longer the intrinsic material property but rather the specific loading condition. Representative examples are Ti$_{5}$Nb at 300~K and Ti$_{4}$Nb at 77~K. This explains why conventional theoretical approaches that rely primarily on $\beta$-phase stability cannot distinguish twinning modes within the TWIP regime. Moreover, both TWIP and TRIP can also be tuned by loading conditions when $P^{\beta \leftrightarrow \alpha''}_{\mathrm{chem}}$ lies between 0 and 1, as exemplified by Ti$_{5}$Nb at 77~K. A quantitative investigation of how loading conditions influence deformation mode selection will be an important topic for future studies. 

Compared to the widely used $\overline{\text{Bo}}$–$\overline{\text{Md}}$ method, $P^{\beta \leftrightarrow \alpha''}_{\mathrm{chem}}$ in our framework more clearly captures the correlation between temperature-dependent $\beta$-phase stability, its reversibility, and deformation mode selection. Taking the alloys in this study as examples, the ($\overline{\text{Bo}}$, $\overline{\text{Md}}$) values for Ti$_{5}$Nb, Ti$_{4}$Nb, and Ti$_{2}$Nb are (2.8415, 2.4431), (2.8518, 2.4424), and (2.8930, 2.3933), respectively. Although these temperature-independent $\overline{\text{Bo}}$–$\overline{\text{Md}}$ parameters are quite similar, their deformation modes differ significantly. In particular, Ti$_{5}$Nb and Ti$_{4}$Nb are located near the TRIP–TWIP boundary in the $\overline{\text{Bo}}$–$\overline{\text{Md}}$ diagram (see Figure~\ref{fig:Experimental}\textbf{B}). However, this TRIP/TWIP boundary in the $\overline{\text{Bo}}$–$\overline{\text{Md}}$ diagram is based on empirical fitting rather than a physically derived criterion, making it ambiguous for data points near the boundary. Furthermore, as summarized in Table~\ref{tab:deformationMode}, the effect of temperature on deformation mode selection cannot be neglected.

Another advantage of this framework over conventional semi-empirical methods is its ability to provide insights into twinning mode selection within the TWIP regime under mechanical loading. Since both twinning modes occur via the same reversible $\beta$–$\alpha''$ phase transformation, the elastic strain free energy associated with twinning nucleation plays a more critical role. This elastic strain energy is strongly correlated with the strain state (or structural distortion) of the $\beta$ phase and the nucleated $\alpha''$ phase. The equilibrium structures of the $\alpha''$ and $\beta$ phases vary with alloy composition, temperature, and loading conditions. Due to the metastability of the $\beta$ phase and its relatively low elastic moduli, it is highly susceptible to lattice distortion under applied loading. Moreover, many metastable $\beta$-Ti alloys contain neutral elements such as Sn and Zr, which do not significantly affect $\beta$ phase stability but can substantially modify lattice parameters, elastic moduli, and elastic constants of both $\beta$ and $\alpha''$ phases, thereby influencing the elastic strain energy penalty for twin nucleation~\cite{kim2015crystal,hao2006effect}. In addition, $\alpha''$-stabilizing elements such as oxygen, which are often introduced during materials processing, can also induce lattice distortions and alter the equilibrium lattice parameters of both $\beta$ and $\alpha''$ phases~\cite{obbard2011effect,chou2019oxygen,nakai2009effect,wang2021roles,tahara2016role}. Therefore, structural distortions and lattice modulations in $\beta$ and $\alpha''$ phases can become significant under mechanical loading but are poorly captured by existing empirical alloy design methods. The strain state of metastable phases is represented by a second-order tensor field, making it difficult to directly link to twin nucleation barriers. In our theoretical framework, $\Delta g^{*}_{\text{twin}}$ effectively connects the strain states of the $\beta$ and $\alpha''$ phases to the misfit strain energy density, thereby determining the competition between different twinning modes. $\Delta g^{*}_{\text{twin}}$ at the undeformed state reveals the intrinsic tendency for twinning mode selection, while $\Delta g^{*}_{\text{twin}}$ at the deformed state illustrates how loading conditions influence twin nucleation behavior and provides guidance for tuning twinning modes through mechanical loading.

Beyond the Ti–Nb and Ti–Mo binary systems included in this study, our deformation mode assessment flowchart is readily extensible to multicomponent metastable $\beta$-Ti alloys, making it a potentially useful tool for alloy design. All these input parameters in this framework of Fig.~\ref{fig:flowchart} can be obtained through both computational and experimental approaches and are not limited to specific alloy systems. Compared with case-by-case MD simulations of complete deformation processes, which require large supercells and entail high computational cost, the parameters needed for our assessment can be obtained much more efficiently. Therefore, this workflow enables high-throughput screening of compositional space to identify alloys that intrinsically favor targeted deformation mechanisms. Furthermore, machine-learning interatomic potentials (MLIPs) trained on high-quality \textit{ab initio} data can further accelerate the prediction of the required energetic and elastic parameters across broad compositional and temperature ranges. Compared with direct atomistic simulations of deformation mechanisms using very large supercells, it is also far more practical to train accurate MLIPs to compute $P^{\beta \leftrightarrow \alpha''}_{\mathrm{chem}}$ and $\Delta g^{*}_{\text{twin}}$ using relatively small cells. Overall, the proposed theoretical framework, when combined with machine-learning techniques, offers a data-efficient prescreening strategy for rapidly identifying alloys with an intrinsic propensity for specific deformation modes~\cite{mishin2021machine,anstine2023machine,mortazavi2023atomistic,freitas2022machine}.

Despite the strengths of the theoretical framework discussed in the previous paragraphs, there are some limitations in its current form. First, $\Delta g^{*}_{\text{twin}}$ in the present framework serves as a metric for estimating trends in nucleation energy barriers across different alloys and conditions, rather than representing a per-event nucleation barrier. As such, it does not explicitly account for the critical nucleation size of twin embryos or the true nucleation barrier that would be obtained through a rigorous variational method. In addition, unlike the assumptions of classical nucleation theory, twin embryos are not spherical but typically plate-, wedge-, or needle-shaped, depending on elastic accommodation under complex loading conditions. Their morphology is determined by elastic accommodation during deformation and can influence the relative nucleation propensities of competing twinning modes. Accurate estimation of the shapes and aspect ratios of twin embryos requires quantitative information on the elastic anisotropy of the alloy, the applied loading conditions, and the thermodynamic driving forces, including chemical free energy, in-plane misfit strain energy, and out-of-plane interfacial free energy. The accuracy of our nucleation metric can be further improved by incorporating elastic anisotropy and out-of-plane interfacial energy density into the current framework ~\cite{liu2023three,khachaturyan2013theory,ledbetter1999habit,lee1990elastic,yoo1991deformation,yang2021332,li2023deformation}.

Another limitation of the current framework is the lack of information on dynamic growth behaviors. In particular, within the TWIP regime, the final volume fraction of twins in the material, which correlates with their contribution to plasticity, is highly dependent on twin growth behavior. For the \{112\} twin in metastable Ti alloys, the growth behavior is closely related to the stability of the $\omega$ phase~\cite{chen2024stability}, and the growth kinetics can be tuned by controlling the reversibility between the $\beta$ and $\omega$ phases. In contrast, \{112\} twin nucleation and growth in stable BCC materials do not necessarily require phase transformations and can instead be triggered by successive slip on \{112\} planes~\cite{li2023deformation}.  Since our focus is on the metastable region, we do not consider this successive slip mechanism here, particularly because dislocation slip is the dominant carrier of plastic deformation compared to \{112\} twinning in stable BCC materials. For the \{332\} twin in metastable Ti alloys, $\alpha''$ phase is often observed along twin boundaries~\cite{lai2016mechanism,chen2018transitional} (including in this study), suggesting twin growth via reversible $\beta$–$\alpha''$ phase transformation. However, the atomistic motion of twin boundaries during growth remains under debate and does not strictly follow the Burgers path~\cite{tobe2014origin,kwasniak2022polymorphic}. To fully evaluate the growth kinetics of \{332\} twins, detailed investigations of atomic motions at twin boundaries are still required in future studies.


In summary, our atomistic simulations capture the full spectrum of common deformation modes in metastable $\beta$-Ti alloys. We then derive a $P^{\beta \leftrightarrow \alpha''}_{\mathrm{chem}}$–$\Delta g^{*}_{\text{twin}}$ flowchart, which incorporates both the free energy barriers for homogeneous transformations and the misfit strain energy along specific phase boundaries, based on structural analyses of simulation data combined with crystallographic theory. This conceptual flowchart provides clear criteria for the nucleation of different deformation modes, with parameters that correlate temperature, alloy composition, and loading conditions during mechanical deformation. In the \textbf{Supplementary Materials}, we also provide a step-by-step workflow for calculating all key parameters in the flowchart. The applicability of this framework is not limited to the alloys studied here; it can be extended to multicomponent metastable $\beta$-Ti alloys and to the future design of metastable BCC refractory high-entropy alloys using high-throughput atomistic simulations of defect-free bulk systems based on accurate MLIPs.

\section*{Materials and Methods}
\label{methods}

\begin{figure}[htbp]
    \centering
    \makebox[\textwidth][c]{%
    \includegraphics[width=1.1\textwidth, clip=true, trim=0 0 0 0]{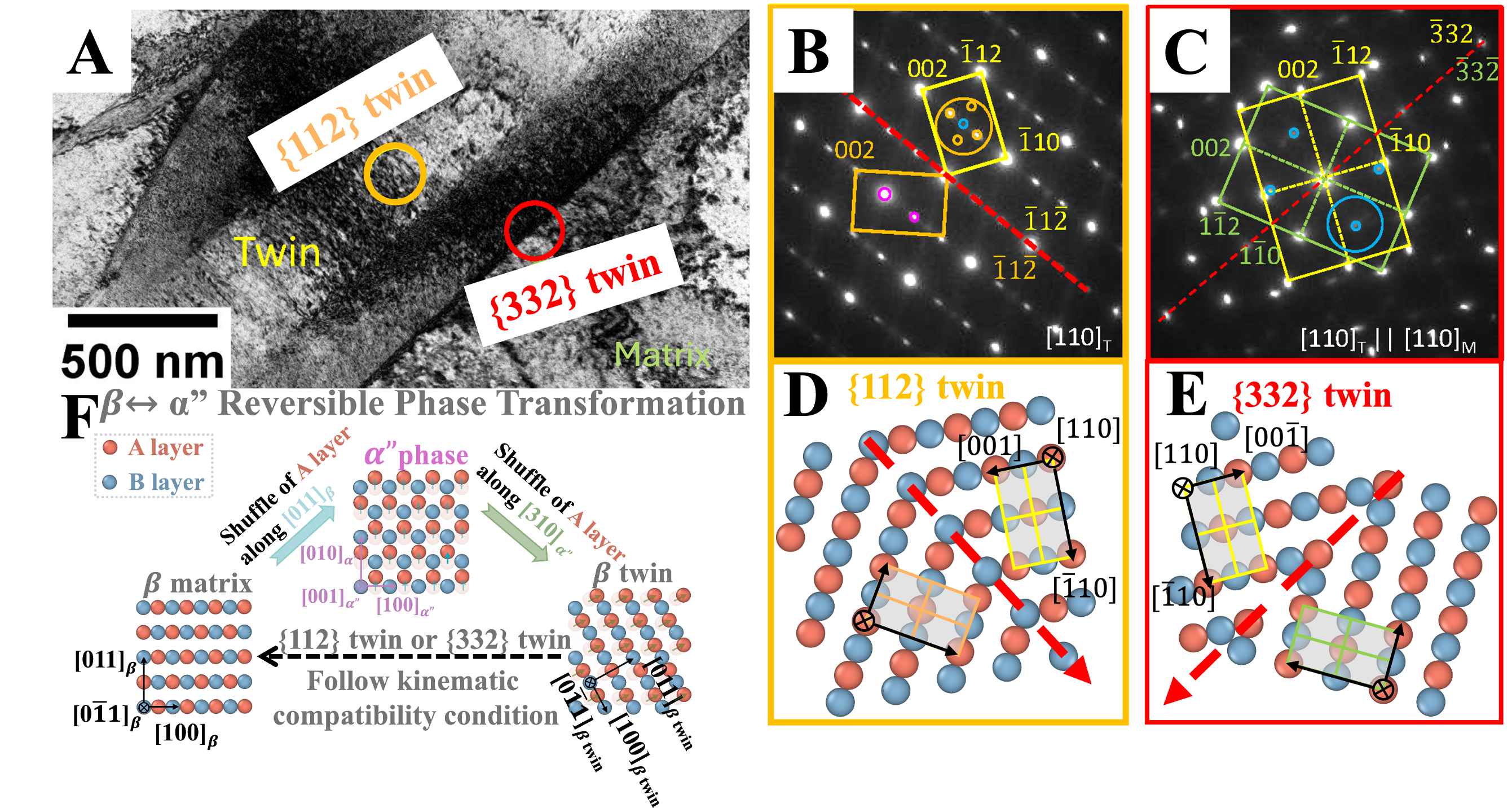}
    }
    \caption{\textbf{Coexistence of \{112\} twin and \{332\} twin in 5$\%$ cold-rolled Ti2448 (Ti–24Nb–4Zr–8Sn) sample.} \textbf{A} is the TEM figure illustrating morphology of \{112\} twin and \{332\} twin, red circle mark the boundary of \{332\} twin, and orange circle mark the boundary of secondary \{112\} twin located inside the \{332\} twin region. \textbf{B-C} are the diffraction patterns at the boundary of\{112\} twin and \{332\} twin. \textbf{D-E} are schematic draws for the \{112\} twin boundary and \{332\} twin boundary.\textbf{F} show an example of formation of BCC twin phase via reversible $\beta$-$\alpha"$ phase transformations.}
    \label{fig:Coexist}
\end{figure}

\paragraph*{Molecular dynamics simulations on nucleation of defects}
\label{simulation set on loading} We use the Large-Scale Atomic/Mol-ecular Massively Parallel Simulator (LAMMPS~\cite{plimpton1995fast,thompson2022lammps}) to perform \ac{MD} simulations. We investigate temperature-, chemical composition- and loading condition- dependent twin nucleation in Ti-Nb binary alloy system using \ac{MEAM} potential developed by Ehemann et al.~\cite{ehemann2017force}. As for the simulation box setting, a supercell containing 405504 atoms with basis vectors of 32[311]$\times$12$[\bar{2}33]$$\times$48[01$\bar{1}$] in the unit of conventional BCC lattice vectors. As for temperature and chemical composition setting, we perform loading operations under 77~K (liquid nitrogen temperature) and 300~K (room temperature), which are two common operation temperatures in experiments. Meanwhile, we selected three different compositions that cross the compositional boundaries between the unstable–metastable and metastable–stable regions of the BCC phase: Ti$_{5}$Nb (Ti-27 wt\%Nb, close to Nb concentration in experimental sample in ~\ref{fig:Coexist}),Ti$_{4}$Nb (Ti-33 wt\%Nb )and Ti$_{2}$Nb (Ti-49 wt\%Nb). As for the loading conditions, we apply shear loading along [311] direction on $(\bar{2}33)$ with two ways of setting the boundary conditions:
The first boundary condition corresponds to a confined simple shear state. By controlling strain condition, only the shear strain is applied and the other strain components should be constrained to 0:
\[
\varepsilon_{ij}^{\text{(confined simple shear)}} =
\begin{bmatrix}
0 & \gamma_{xy} & 0 \\
0 & 0 & 0 \\
0 & 0 & 0
\end{bmatrix}
\quad \text{in basis: }
X = [3\,1\,1], \quad
Y = [\overline{2}\,3\,3], \quad
Z = [0\,1\,\overline{1}]
\]
The second setting on boundary condition is to apply non-confined simple shear for the system. By controlling stress condition, only shear stress is applied and the other non-shear stress components should relax to 0 and in a symmetric Cauchy stress tensor form:
\[
\sigma_{ij}^{\text{(non-confined simple shear)}} =
\begin{bmatrix}
0 & \tau_{xy} & 0 \\
\tau_{yx} & 0 & 0 \\
0 & 0 & 0
\end{bmatrix}
\quad \text{in basis: }
X = [3\,1\,1], \quad
Y = [\overline{2}\,3\,3], \quad
Z = [0\,1\,\overline{1}]
\]
Before dynamic loading, we hold the simulation box under isothermal-isobaric (NPT) Nose-Hoover dynamics\cite{melchionna1993hoover} ($P$ = 0, and $T$ is consistent with further loading temperature) for 50 ps to achieve thermal equilibrium. The timestep is set to 1 fs and the engineering shear strain rate is \( 1 \times 10^8 \)/s.

\paragraph*{Metadynamics for phase transformation barriers }
\label{Metadynamics}
To explore the free energy landscape associated with the BCC–$\alpha"$ phase transformation, we employ well-tempered metadynamics using the PLUMED plugin ~\cite{tribello2014plumed,plumed2019promoting}.The whole simulation box is under isothermal–isobaric conditions using Parrinello–Rahman–type barostat (NPH) combined with canonical sampling via velocity rescaling with Hamiltonian dynamics (temp/csvr) at target temperature~\cite{shinoda2004rapid,bussi2007canonical}. Here we do not use conventional NPT method to simulate isothermal–isobaric conditions since NPH$+$temp/csvr can be more reliably to control the system temperature during bias deposition at non-equilibrium state. The orientation of simulation box along the BCC crystal directions: \(8[100] \times 6[011] \times 6[0\bar{1}1]\), with a total of 1152 atoms. Here the supercell size is smaller than the supercell size (18000 atoms) for lattice constants calculation. This choice ensures that the entire simulation cell can undergo a homogeneous lattice transformation along the Burgers pathway, enabling efficient exploration of the free energy surface. Larger systems are not suitable for this type of metadynamics study because they introduce heterogeneous nucleation and significantly increase the computational cost, without affecting the intrinsic homogeneous chemical free energy barrier we aim to quantify. Two collective variables (CVs) are chosen to follow Burgers ~\cite{burgers1934process} transformation mechanisms: (i) the relative shuffle displacement between adjacent \{110\} planes, computed as the $z$-component of the center-of-mass distance between two groups of atoms on alternative \{110\} planes, and (ii) the axial strain represented by the $b/a$ ratio, evaluated as the ratio of the simulation cell lengths along the $[011]$ and $[100]$ directions, respectively.

The metadynamics bias is applied using Gaussian hills of height 0.05 meV/atom, deposited every 200 steps. The width (sigma) of the Gaussians for each CV was set to 5\% of its typical fluctuation range to ensure resolution without over-biasing. A bias factor of 30 is used to modulate the bias deposition in accordance with the well-tempered metadynamics protocol, enabling efficient exploration of metastable states while avoiding oversampling of already visited configurations. All the metadynamics simulations are performed for a total of 100 nanoseconds to ensure sufficient sampling and convergence of the free energy surface.

\paragraph*{Elastic constants on metastable phases }
\label{Elastic constants}
To determine the elastic constants of the body-centered cubic (BCC) phase, we employed the finite deformation (perturbation) method ~\cite{ray1984statistical,zhen2012deformation} using LAMMPS. This approach involves applying small homogeneous deformations to the equilibrium structure and measuring the resulting stress response. For BCC systems, only three independent elastic constants exist: \( C_{11} \), \( C_{12} \), and \( C_{44} \). We apply the following deformation modes:

\begin{itemize}
  \item Uniaxial strain along the $x$-axis: \( \varepsilon_{xx} = \delta \) to compute \( C_{11} \) from \( \sigma_{xx} \).
  \item Volume-conserving orthorhombic strain: \( \varepsilon_{xx} = \delta, \, \varepsilon_{yy} = -\delta \) to compute \( C_{11} - C_{12} \) from \( \sigma_{xx} - \sigma_{yy} \).
  \item Simple shear: \( \varepsilon_{xy} = \delta \) to obtain \( C_{44} \) from \( \sigma_{xy} \).
\end{itemize}
In each case, small strain magnitudes ( \( \delta = \pm 0.001 \)) are used to ensure linear response. The stress–strain relationship follows Hooke’s law and from which elastic constants are extracted by linear fitting. All simulations are conducted at 0~K to eliminate thermal noise and isolate the mechanical response~\cite{nye1985physical}.

\paragraph{Directional elastic moduli and Poisson's ratio}
Here we provide the general formula of directional Young's modulus, directional shear modulus and directional Poisson's ratio for cubic systems.

\textbf{Directional Shear modulus}
To compute the shear modulus \( G \) along an arbitrary slip system in a cubic crystal, one must consider both the shear direction \( \mathbf{d} = (d_1, d_2, d_3) \) and the slip plane normal \( \mathbf{n} = (n_1, n_2, n_3) \), expressed in the crystal coordinate system. The general form of the inverse shear modulus for cubic system based on the elastic compliance tensor \( S_{ijkl} \)  can be expressed explicitly as:
\begin{equation}
\begin{aligned}
\frac{1}{G(\bn)} =\ & 4 \left\{ S_{11} \left[ (d_1 n_1)^2 + (d_2 n_2)^2 + (d_3 n_3)^2 \right] 
+ 2 S_{12} (d_1 n_1 d_2 n_2 + d_1 n_1 d_3 n_3 + d_2 n_2 d_3 n_3) \right\} \\
& + S_{44} \left[ (d_2 n_3 + d_3 n_2)^2 + (d_1 n_3 + d_3 n_1)^2 + (d_1 n_2 + d_2 n_1)^2 \right]
\end{aligned}
\label{eq:GeneralShearModulus}
\end{equation}

\textbf{Directional Young's modulus}
Let $\bn=(l_1,l_2,l_3)$ be a unit vector giving the loading direction
Apply a uniaxial stress $\sigma$ along $\bn$, then the general form of the inverse Young's modulus in cubic system based on the elastic compliance tensor \( S_{ijkl} \) is:
\begin{equation}
\begin{aligned}
\frac{1}{E(\bn)}
=
S_{11}\big(l_1^4+l_2^4+l_3^4\big)
+2S_{12}\big(l_1^2l_2^2+l_2^2l_3^2+l_3^2l_1^2\big)
+S_{44}\big(l_1^2l_2^2+l_2^2l_3^2+l_3^2l_1^2\big).
\end{aligned}
\label{eq:GeneralYoung"sModulus}
\end{equation}

\textbf{Directional Poisson's ratio}
As for the derivation for general form of directional Poisson's ratio, here we define two unit vector: \( \mathbf{p} = (p_1, p_2, p_3) \) for longitudinal direction and \( \mathbf{q} = (q_1, q_2, q_3) \) for transverse direction, and then it can be expressed explicitly in cubic system as
\begin{equation}
\begin{aligned}
v_{a\rightarrow b} &=-\frac{\varepsilon_{bb}}{\varepsilon_{aa}}=  -\frac{S_{12}+(S_{11}-S_{12}-\frac{1}{2}S_{44})(p_1^2 q_1^2 + p_2^2 q_2^2 + p_3^2 q_3^2)} {S_{11}-2(S_{11}-S_{12}-\frac{1}{2}S_{44})(p_1^2p_2^2 + p_2^2p_3^2 + p_3^2p_1^2)}  \\
\end{aligned}
\label{eq:GeneralPossion"s}
\end{equation}
\paragraph*{Molecular dynamics simulations on calculation of lattice constants at undeformed state}
\label{Lattice Constants}
To calculate lattice constants of BCC phase and $\alpha"$ phase at stress-free condition under finite temperatures, we construct supercells in perfect BCC and $\alpha"$ phase and then relax at finite temperatures. As for the supercell in perfect BCC phase, the basis is 20[100]$\times$15[011]$\times$15$[0\bar{1}1]$ in the unit of conventional BCC lattice vectors. As for supercell in perfect HCP phase, the basis is 20/3$[11\bar{2}0]$$\times$15$[\bar{1}100]$$\times$15$[0001]$. We then relax the supercells at finite temperature (77~K or 300~K) and hold the simulation box under isothermal-isobaric (NPT) Nose-Hoover dynamics\cite{melchionna1993hoover} ($P$ = 0) for 100 ps to achieve thermal equilibrium. Then we average the lattice parameters of last 100 timesteps. Since at metastable BCC region, BCC phase can locally transfer to other metastable phases ($\omega$ or $\alpha"$ phase) and so we first use PTM methods to filter out the atoms in BCC phase. Then we calculate the unit length along X=[100],Y=[011] and Z=$[0\bar{1}1]$ within BCC phase. Here unit length along [011]/$[0\bar{1}1]$ is not rigorously equal to $\sqrt{2}$ a$_{\text{$\beta$}}$(unit length of [100]) when BCC phase is metastable in $\beta$ Ti alloys.  Finally we can obtain average lattice constant $a_{\text{$\beta$}}$ of BCC phases at equilibrium state. As for lattice constants of $\alpha"$ phase, once the supercells in HCP phase is relaxed at finit temperature, the simulation box will spontaneously transfer to metastable $\alpha"$ phase. Similar to the case of BCC phase, we first use PTM method to extract atoms in $\alpha"$ phase. Then we calculate unit length along $[11\bar{2}0]_{\text{HCP}}$$\parallel$$[100]_{\text{$\alpha"$}}$,$[\bar{1}100]_{\text{HCP}}$$\parallel$$[010]_{\text{$\alpha"$}}$ and $[0001]_{\text{HCP}}$$\parallel$$[001]_{\text{$\alpha"$}}$. The unit lengths of corresponding directions are length of $a_{\text{$\alpha"$}}$, $b_{\text{$\alpha"$}}$ and $c_{\text{$\alpha"$}}$, respectively.

\paragraph{Misfit strain along interfaces between deformed BCC and $\alpha"$ phase}
To calculate misfit strain along interphase boundaries at deformed state, we need information on deform state of $\beta$ parent phase and nucleated $\alpha"$ phase. Since under deformed state, the grains undergo stretch and rotation, the structure of phases are deviated from stress-free state and we cannot directly calculate needed lattice constants. The information on deformed state of lattices of $\beta$ and $\alpha"$ phase can be extracted from elastic deformation gradient matrix obtained from PTM method from OVITO. In brief, we first choose the reference state for BCC phase and $\alpha"$ phase and define corresponding deformed state under loading conditions. Both reference state and deformed state are described by elastic deformation gradient matrix at the global reference of perfect BCC and HCP unit cell defined in OVITO ~\cite{larsen2016robust}. Further, elastic deformation gradient matrix of deformed state relative to reference state we choose will be then used for misfit strain calculations. Here we can perform polar decomposition on the elastic deformation gradient matrix to obtain pure stretch part and pure rotation part:
\begin{equation}
\begin{aligned}
\mathbf{F} = \mathbf{R} \cdot \mathbf{U}
\end{aligned}
\label{PolarDecomposition}
\end{equation}
Here $\mathbf{U}$ is the right stretch tensor (symmetric and positive-definite), and $\mathbf{R}$ is the rotation matrix (proper orthogonal). 
And right stretch tensor $\mathbf{U}$ can be extracted as:
\begin{equation}
\mathbf{U} = \sqrt{\mathbf{F}^\mathrm{T} \cdot \mathbf{F}}
\label{RightStretchTensor}
\end{equation}
Elastic stretch \textbf{U} here only characterize the shape of current deformed state relative to reference state. For example, if the lattice isotopically expand 2$\%$ while relative ratio of unit length of coordination system is unchanged compared to reference state, the elastic stretch tensor will remain in the identity matrix. Hence if we want to obtain the real unit length along different directions, we need to transfer the elastic stretch tensor to real stretch tensor by multiplying the lattice constants at deformed state. In ~\textbf{Supplementary Materials}, we present a step-by-step procedures on how to calculate misfit strain along two interphase boundaries defined in Equation ~\ref{eq:OR_variant} from information on deformed state of lattice and lattice constants under deformed state.

\newpage


\begin{table}[htbp]
\centering
\caption{Deformation mechanisms of Ti-Nb alloys at different temperatures. (shear strain rate is $10^{8}/s$), Here "SS" in the table is short for "Simple Shear".}
\begin{tabular}{@{}lccccc@{}}
\toprule
 & \multicolumn{2}{c}{77~K} & \multicolumn{2}{c}{300~K} & \multirow{2}{*}{\makecell[c]{Predicted by\\ $\overline{\mathrm{Bo}}$--$\overline{\mathrm{Md}}$}} \\
\cmidrule(lr){2-3} \cmidrule(lr){4-5}
Alloy & Confined SS & Non-confined SS & Confined SS & Non-confined SS \\
\midrule
Ti$_{5}$Nb & \{332\} twin & $\alpha''$+ local \{112\} twin & \{332\} twin & $\alpha''$ + local \{112\} twin & TRIP\\
Ti$_{4}$Nb & \{332\} twin & \{112\} twin & \{112\} twin & \{112\} twin & TWIP \\
Ti$_{2}$Nb & Dislocation & \{112\} twin + Dislocation & Dislocation & Dislocation & Dislocation\\
\bottomrule
\end{tabular}
\label{tab:deformationMode}
\end{table}

\begin{table}[htbp]
\centering
\caption{Summary on the energetic descriptors for twinning nucleation behaviors and reversibility between $\beta$ and $\alpha"$ phase. Unit of all energetic parameters is meV/atom. $P^{\mathrm{\beta}\leftrightarrow\alpha''}_{\mathrm{chem}}$ is dimensionless. }
\small
\setlength{\tabcolsep}{1pt} 
\renewcommand{\arraystretch}{1.12}
\begin{tabular*}{\linewidth}{@{}
l
S[table-format=1.2]
S[table-format=1.3]
S[table-format=1.3]
S[table-format=2.2]
S[table-format=1.2]
S[table-format=1.2]
p{55mm}%
@{}} 
\toprule
\textbf{Alloy System} &
{$\left| \Delta F^{\textup{A}}_{\beta \leftrightarrow \alpha''} \right|$} &
{$w^{\mathrm{misfit}}_{\mathrm{\{112\}twin}}$} &
{$w^{\mathrm{misfit}}_{\mathrm{\{332\}twin}}$} &
{$\mathbf{P^{\mathrm{\beta}\leftrightarrow\alpha''}_{\mathrm{chem}}}$} &
{$\Delta g^{*}_{\text{\{112\}twin}}$} & 
{$\Delta g^{*}_{\text{\{332\}twin}}$} & 
\textbf{Deformation Modes }\\
\midrule
Ti$_5$Nb@77~K(MD)  & 2.76 & 5.33 & 5.23 & \textbf{0.66} & \textbf{8.09}  & \textbf{7.99}  & \dm{$\alpha"$/\{332\} twin}\\
Ti$_5$Nb@300~K(MD)  & 3.52 & 3.16 & 6.74 & \textbf{0.87} & \textbf{6.68}  & \textbf{10.26}  &\dm{\{332\} twin/\{112\} twin}\\
Ti$_4$Nb@77~K(MD)   & 1.35 & 3.22 & 5.74 & \textbf{0.82} & \textbf{4.57}  & \textbf{7.09}  &\dm{\{332\} twin/\{112\} twin}\\
Ti$_4$Nb@300~K(MD)  & 2.27 & 1.84 & 6.81 & \textbf{0.92} & \textbf{4.10}  & \textbf{9.08}  &\dm{\{112\} twin}\\
Ti10Mo@300~K(Exp.)  & 2.75 & 8.88 & 0.30 & \textbf{0.90} & \textbf{11.63} & \textbf{3.05}  &\dm{\{332\} twin}\\
\bottomrule
\end{tabular*}
\label{tab:StrainCalc}
\end{table}


\clearpage 

%
\bibliography{main} 

\begin{thebibliography}{100}
\providecommand{\url}[1]{\texttt{#1}}
\expandafter\ifx\csname urlstyle\endcsname\relax
  \providecommand{\doi}[1]{doi:\discretionary{}{}{}#1}\else
  \providecommand{\doi}{doi:\discretionary{}{}{}\begingroup \urlstyle{rm}\Url}\fi

\bibitem{raabe2019metastability}
D.~Raabe, Z.~Li, D.~Ponge, Metastability alloy design. \emph{MRS Bulletin} \textbf{44}~(4), 266--272 (2019).

\bibitem{kolli2018review}
R.~P. Kolli, A.~Devaraj, A review of metastable beta titanium alloys. \emph{Metals} \textbf{8}~(7), 506 (2018).

\bibitem{pesode2023review}
P.~Pesode, S.~Barve, A review—metastable $\beta$ titanium alloy for biomedical applications. \emph{Journal of Engineering and Applied Science} \textbf{70}~(1), 25 (2023).

\bibitem{khachaturyan2013theory}
A.~G. Khachaturyan, \emph{Theory of structural transformations in solids} (Courier Corporation) (2013).

\bibitem{christian1975theory}
J.~W. Christian, The Theory of Transformations in Metals and Alloys. I. \emph{Equilibrium and general kinetic theory.}  (1975).

\bibitem{christian1995deformation}
J.~W. Christian, S.~Mahajan, Deformation twinning. \emph{Progress in materials science} \textbf{39}~(1-2), 1--157 (1995).

\bibitem{saito2003multifunctional}
T.~Saito, \emph{et~al.}, Multifunctional alloys obtained via a dislocation-free plastic deformation mechanism. \emph{Science} \textbf{300}~(5618), 464--467 (2003).

\bibitem{bouaziz2011high}
O.~Bouaziz, S.~Allain, C.~Scott, P.~Cugy, D.~Barbier, High manganese austenitic twinning induced plasticity steels: A review of the microstructure properties relationships. \emph{Current opinion in solid state and materials science} \textbf{15}~(4), 141--168 (2011).

\bibitem{de2018twinning}
B.~C. De~Cooman, Y.~Estrin, S.~K. Kim, Twinning-induced plasticity (TWIP) steels. \emph{Acta Materialia} \textbf{142}, 283--362 (2018).

\bibitem{fischer1996transformation}
F.~Fischer, Q.~Sun, K.~Tanaka, Transformation-induced plasticity (TRIP). \emph{Applied Mechanics Reviews} \textbf{49}~(6), 317 (1996).

\bibitem{fischer2000new}
F.-D. Fischer, \emph{et~al.}, A new view on transformation induced plasticity (TRIP). \emph{International Journal of Plasticity} \textbf{16}~(7-8), 723--748 (2000).

\bibitem{olson1978transformation}
G.~B. Olson, M.~Azrin, Transformation behavior of TRIP steels. \emph{Metallurgical Transactions A} \textbf{9}, 713--721 (1978).

\bibitem{banerjee2013perspectives}
D.~Banerjee, J.~Williams, Perspectives on titanium science and technology. \emph{Acta Materialia} \textbf{61}~(3), 844--879 (2013).

\bibitem{boyer1996overview}
R.~R. Boyer, An overview on the use of titanium in the aerospace industry. \emph{Materials Science and Engineering: A} \textbf{213}~(1-2), 103--114 (1996).

\bibitem{brozek2016beta}
C.~Brozek, \emph{et~al.}, A $\beta$-titanium alloy with extra high strain-hardening rate: design and mechanical properties. \emph{Scripta Materialia} \textbf{114}, 60--64 (2016).

\bibitem{leyens2006titanium}
C.~Leyens, M.~Peters, \emph{Titanium and titanium alloys: fundamentals and applications} (Wiley Online Library) (2006).

\bibitem{niinomi2008mechanical}
M.~Niinomi, Mechanical biocompatibilities of titanium alloys for biomedical applications. \emph{Journal of the mechanical behavior of biomedical materials} \textbf{1}~(1), 30--42 (2008).

\bibitem{marteleur2012design}
M.~Marteleur, \emph{et~al.}, On the design of new $\beta$-metastable titanium alloys with improved work hardening rate thanks to simultaneous TRIP and TWIP effects. \emph{Scripta Materialia} \textbf{66}~(10), 749--752 (2012).

\bibitem{hao2016superelasticity}
Y.~Hao, \emph{et~al.}, Superelasticity and tunable thermal expansion across a wide temperature range. \emph{Journal of Materials Science \& Technology} \textbf{32}~(8), 705--709 (2016).

\bibitem{li2023deformation}
X.~Li, Z.~Zhang, J.~Wang, Deformation twinning in body-centered cubic metals and alloys. \emph{Progress in Materials Science} \textbf{139}, 101160 (2023).

\bibitem{wang2024tizrhfnb}
S.~Wang, \emph{et~al.}, TiZrHfNb refractory high-entropy alloys with twinning-induced plasticity. \emph{Journal of Materials Science \& Technology} \textbf{187}, 72--85 (2024).

\bibitem{lilensten2017design}
L.~Lilensten, \emph{et~al.}, Design and tensile properties of a bcc Ti-rich high-entropy alloy with transformation-induced plasticity. \emph{Materials Research Letters} \textbf{5}~(2), 110--116 (2017).

\bibitem{huang2017phase}
H.~Huang, \emph{et~al.}, Phase-transformation ductilization of brittle high-entropy alloys via metastability engineering. \emph{Advanced Materials} \textbf{29}~(30), 1701678 (2017).

\bibitem{wang2019formation}
L.~Wang, \emph{et~al.}, Formation and toughening of metastable phases in TiZrHfAlNb medium entropy alloys. \emph{Materials Science and Engineering: A} \textbf{748}, 441--452 (2019).

\bibitem{zhang2018phase}
L.~Zhang, \emph{et~al.}, Phase transformations in body-centered cubic NbxHfZrTi high-entropy alloys. \emph{Materials Characterization} \textbf{142}, 443--448 (2018).

\bibitem{wang2020mechanical}
S.~Wang, \emph{et~al.}, Mechanical instability and tensile properties of TiZrHfNbTa high entropy alloy at cryogenic temperatures. \emph{Acta Materialia} \textbf{201}, 517--527 (2020).

\bibitem{cao2024review}
M.~Cao, B.~He, A review on deformation mechanisms of metastable $\beta$ titanium alloys. \emph{Journal of Materials Science} \textbf{59}~(32), 14981--15016 (2024).

\bibitem{gao2018deformation}
J.~Gao, \emph{et~al.}, Deformation mechanisms in a metastable beta titanium twinning induced plasticity alloy with high yield strength and high strain hardening rate. \emph{Acta Materialia} \textbf{152}, 301--314 (2018).

\bibitem{hanada1987correlation}
S.~Hanada, O.~Izumi, Correlation of tensile properties, deformation modes, and phase stability in commercial $\beta$-phase titanium alloys. \emph{Metallurgical and Materials Transactions A} \textbf{18}, 265--271 (1987).

\bibitem{min2013mechanism}
X.~Min, X.~Chen, S.~Emura, K.~Tsuchiya, Mechanism of twinning-induced plasticity in $\beta$-type Ti--15Mo alloy. \emph{Scripta Materialia} \textbf{69}~(5), 393--396 (2013).

\bibitem{sun2013investigation}
F.~Sun, \emph{et~al.}, Investigation of early stage deformation mechanisms in a metastable $\beta$ titanium alloy showing combined twinning-induced plasticity and transformation-induced plasticity effects. \emph{Acta Materialia} \textbf{61}~(17), 6406--6417 (2013).

\bibitem{blackburn1971stress}
M.~Blackburn, Stress-induced transformations in Ti--Mo alloys. \emph{J. Inst. Metals} \textbf{99}, 132 (1971).

\bibitem{zhang2019strong}
J.~Zhang, \emph{et~al.}, Strong and ductile beta Ti--18Zr--13Mo alloy with multimodal twinning. \emph{Materials research letters} \textbf{7}~(6), 251--257 (2019).

\bibitem{gao2016group}
Y.~Gao, R.~Shi, J.-F. Nie, S.~A. Dregia, Y.~Wang, Group theory description of transformation pathway degeneracy in structural phase transformations. \emph{Acta Materialia} \textbf{109}, 353--363 (2016).

\bibitem{gao2020intrinsic}
Y.~Gao, Y.~Zheng, H.~Fraser, Y.~Wang, Intrinsic coupling between twinning plasticity and transformation plasticity in metastable $\beta$ Ti-alloys: a symmetry and pathway analysis. \emph{Acta Materialia} \textbf{196}, 488--504 (2020).

\bibitem{groger2023twinning}
R.~Gr{\"o}ger, J.~Holzer, T.~Kruml, Twinning and antitwinning in body-centered cubic metals. \emph{Computational Materials Science} \textbf{216}, 111874 (2023).

\bibitem{antonov2020novel}
S.~Antonov, \emph{et~al.}, Novel deformation twinning system in a cold rolled high-strength metastable-$\beta$ Ti-5Al-5V-5Mo-3Cr-0.5 Fe alloy. \emph{Materialia} \textbf{9}, 100614 (2020).

\bibitem{crocker1962twinned}
A.~Crocker, Twinned martensite. \emph{Acta Metallurgica} \textbf{10}~(2), 113--122 (1962).

\bibitem{tobe2014origin}
H.~Tobe, H.~Y. Kim, T.~Inamura, H.~Hosoda, S.~Miyazaki, Origin of $\{$3 3 2$\}$ twinning in metastable $\beta$-Ti alloys. \emph{Acta materialia} \textbf{64}, 345--355 (2014).

\bibitem{hanada1985deformation}
S.~Hanada, M.~Ozeki, O.~Izumi, Deformation characteristics in $\beta$ phase Ti-Nb alloys. \emph{Metallurgical transactions A} \textbf{16}, 789--795 (1985).

\bibitem{zhang2020hierarchical}
J.~Zhang, \emph{et~al.}, Hierarchical \{332\}$\langle$113$\rangle$ twinning in a metastable $\beta$ Ti-alloy showing tolerance to strain localization. \emph{Materials Research Letters} \textbf{8}~(7), 247--253 (2020).

\bibitem{chen2018transitional}
B.~Chen, W.~Sun, Transitional structure of \{332\}$\langle$113$\rangle$$\beta$ twin boundary in a deformed metastable $\beta$-type Ti-Nb-based alloy, revealed by atomic resolution electron microscopy. \emph{Scripta Materialia} \textbf{150}, 115--119 (2018).

\bibitem{lai2016mechanism}
M.~Lai, C.~C. Tasan, D.~Raabe, On the mechanism of $\{$332$\}$ twinning in metastable $\beta$ titanium alloys. \emph{Acta Materialia} \textbf{111}, 173--186 (2016).

\bibitem{oka1978stress}
M.~Oka, Y.~Taniguchi, Stress-Induced Products in Metastable Beta Ti--Mo and Ti--V Alloys. \emph{Journal of the Japan Institute of Metals} \textbf{42}~(8), 814--820 (1978).

\bibitem{oka1979332}
M.~Oka, Y.~Taniguchi, $\{$332$\}$ Deformation twins in a Ti-15.5 pct V alloy. \emph{Metallurgical Transactions A} \textbf{10}, 651--653 (1979).

\bibitem{hanada1986transmission}
S.~Hanada, O.~Izumi, Transmission electron microscopic observations of mechanical twinning in metastable beta titanium alloys. \emph{Metallurgical Transactions A} \textbf{17}, 1409--1420 (1986).

\bibitem{bhattacharya2003microstructure}
K.~Bhattacharya, \emph{Microstructure of martensite: why it forms and how it gives rise to the shape-memory effect}, vol.~2 (Oxford University Press) (2003).

\bibitem{wu20141}
S.~Wu, \emph{et~al.}, $\{$112$\}$$\langle$111$\rangle$ Twinning during $\omega$ to body-centered cubic transition. \emph{Acta materialia} \textbf{62}, 122--128 (2014).

\bibitem{chen2024stability}
G.~Chen, D.~Li, Y.~Zheng, L.~Qi, Stability and growth kinetics of $\{$112$\}$ twin embryos in $\beta$-Ti alloys. \emph{Acta Materialia} \textbf{263}, 119520 (2024).

\bibitem{cho2020study}
K.~Cho, R.~Morioka, S.~Harjo, T.~Kawasaki, H.~Y. Yasuda, Study on formation mechanism of $\{$332$\}$$\langle$113$\rangle$ deformation twinning in metastable $\beta$-type Ti alloy focusing on stress-induced $\alpha$” martensite phase. \emph{Scripta Materialia} \textbf{177}, 106--111 (2020).

\bibitem{takemoto1993structural}
Y.~Takemoto, M.~Hida, A.~Sakakibara, Structural relaxation of interface of $\{$332$\}$$\langle$113$\rangle$ twin in $\beta$Ti alloy. \emph{Nippon Kinzoku Gakkaishi (1952)} \textbf{57}~(12), 1471--1472 (1993).

\bibitem{zheng2016effect}
Y.~Zheng, \emph{et~al.}, The effect of alloy composition on instabilities in the $\beta$ phase of titanium alloys. \emph{Scripta Materialia} \textbf{116}, 49--52 (2016).

\bibitem{castany2016reversion}
P.~Castany, Y.~Yang, E.~Bertrand, T.~Gloriant, Reversion of a Parent $\{$130$\}$$\langle$310$\rangle$ $\alpha$" Martensitic Twinning System at the Origin of $\{$332$\}$langle 113rangle $\beta$ Twins Observed in Metastable $\beta$ Titanium Alloys. \emph{Physical review letters} \textbf{117}~(24), 245501 (2016).

\bibitem{hanada1986effect}
S.~Hanada, T.~Yoshio, O.~Izumi, Effect of plastic deformation modes on tensile properties of beta titanium alloys. \emph{Transactions of the Japan institute of metals} \textbf{27}~(7), 496--503 (1986).

\bibitem{zhao2021materials}
G.~Zhao, X.~Li, N.~Petrinic, Materials information and mechanical response of TRIP/TWIP Ti alloys. \emph{npj Computational Materials} \textbf{7}~(1), 91 (2021).

\bibitem{mizutani2010hume}
U.~Mizutani, The Hume-Rothery rules for structurally complex alloy phases, in \emph{Surface properties and engineering of complex intermetallics} (World Scientific), pp. 323--399 (2010).

\bibitem{morinaga1988theoretical}
M.~Morinaga, N.~Yukawa, T.~Maya, K.~Sone, H.~Adachi, Theoretical design of titanium alloys, in \emph{Sixth world conference on titanium. III} (1988), pp. 1601--1606.

\bibitem{coffigniez2024combination}
M.~Coffigniez, \emph{et~al.}, Combination of ab initio descriptors and machine learning approach for the prediction of the plasticity mechanisms in $\beta$-metastable Ti alloys. \emph{Materials \& Design} \textbf{239}, 112801 (2024).

\bibitem{lv2025novel}
M.~Lv, X.~Min, F.~Liu, A novel approach to control the thermal/stress-induced products of body-centered cubic titanium alloys in terms of specific orientation moduli. \emph{Acta Materialia} \textbf{284}, 120594 (2025).

\bibitem{ehemann2017force}
R.~C. Ehemann, J.~W. Wilkins, Force-matched empirical potential for martensitic transitions and plastic deformation in Ti-Nb alloys. \emph{Physical Review B} \textbf{96}~(18), 184105 (2017).

\bibitem{liang2020role}
Q.~Liang, \emph{et~al.}, The role of nano-scaled structural non-uniformities on deformation twinning and stress-induced transformation in a cold rolled multifunctional $\beta$-titanium alloy. \emph{Scripta Materialia} \textbf{177}, 181--185 (2020).

\bibitem{yang2010evolution}
Y.~Yang, \emph{et~al.}, Evolution of deformation mechanisms of Ti--22.4 Nb--0.73 Ta--2Zr--1.34 O alloy during straining. \emph{Acta Materialia} \textbf{58}~(7), 2778--2787 (2010).

\bibitem{zhan2016dynamic}
H.~Zhan, G.~Wang, D.~Kent, M.~Dargusch, The dynamic response of a metastable $\beta$ Ti--Nb alloy to high strain rates at room and elevated temperatures. \emph{Acta Materialia} \textbf{105}, 104--113 (2016).

\bibitem{shin2017phase}
S.~Shin, C.~Zhang, K.~S. Vecchio, Phase stability dependence of deformation mode correlated mechanical properties and elastic properties in Ti-Nb gum metal. \emph{Materials Science and Engineering: A} \textbf{702}, 173--183 (2017).

\bibitem{zhang2019plastic}
W.-d. Zhang, \emph{et~al.}, Plastic deformation mechanism of Ti--Nb--Ta--Zr--O alloy at cryogenic temperatures. \emph{Materials Science and Engineering: A} \textbf{765}, 138293 (2019).

\bibitem{gordin2020new}
D.~Gordin, F.~Sun, D.~Laill{\'e}, F.~Prima, T.~Gloriant, How a new strain transformable titanium-based biomedical alloy can be designed for balloon expendable stents. \emph{Materialia} \textbf{10}, 100638 (2020).

\bibitem{morinaga2018molecular}
M.~Morinaga, The molecular orbital approach and its application to biomedical titanium alloy design, in \emph{Titanium in medical and dental applications} (Elsevier), pp. 39--64 (2018).

\bibitem{plumed2019promoting}
T.~P. consortium, Promoting transparency and reproducibility in enhanced molecular simulations. \emph{Nature methods} \textbf{16}~(8), 670--673 (2019).

\bibitem{tribello2014plumed}
G.~A. Tribello, M.~Bonomi, D.~Branduardi, C.~Camilloni, G.~Bussi, PLUMED 2: New feathers for an old bird. \emph{Computer physics communications} \textbf{185}~(2), 604--613 (2014).

\bibitem{bonomi2009plumed}
M.~Bonomi, \emph{et~al.}, PLUMED: A portable plugin for free-energy calculations with molecular dynamics. \emph{Computer Physics Communications} \textbf{180}~(10), 1961--1972 (2009).

\bibitem{burgers1934process}
W.~Burgers, On the process of transition of the cubic-body-centered modification into the hexagonal-close-packed modification of zirconium. \emph{Physica} \textbf{1}~(7-12), 561--586 (1934).

\bibitem{nishiyama2012martensitic}
Z.~Nishiyama, \emph{Martensitic transformation} (Elsevier) (2012).

\bibitem{zhang2016faceted}
W.-Z. Zhang, X.-F. Gu, F.-Z. Dai, Faceted interfaces: a key feature to quantitative understanding of transformation morphology. \emph{npj Computational Materials} \textbf{2}~(1), 1--14 (2016).

\bibitem{wang2023coherent}
S.~Wang, T.~Wen, J.~Han, D.~J. Srolovitz, Coherent and semicoherent $\alpha$/$\beta$ interfaces in titanium: structure, thermodynamics, migration. \emph{npj Computational Materials} \textbf{9}~(1), 216 (2023).

\bibitem{murzinova2021effect}
M.~Murzinova, S.~Zherebtsov, D.~Klimenko, S.~Semiatin, The effect of $\beta$ stabilizers on the structure and energy of $\alpha$/$\beta$ interfaces in titanium alloys. \emph{Metallurgical and Materials Transactions A} \textbf{52}~(5), 1689--1698 (2021).

\bibitem{shi2012predicting}
R.~Shi, N.~Ma, Y.~Wang, Predicting equilibrium shape of precipitates as function of coherency state. \emph{Acta Materialia} \textbf{60}~(10), 4172--4184 (2012).

\bibitem{jones2018mechanics}
R.~M. Jones, \emph{Mechanics of composite materials} (CRC press) (2018).

\bibitem{eshelby1959elastic}
J.~D. Eshelby, The elastic field outside an ellipsoidal inclusion. \emph{Proceedings of the royal society of London. Series A. Mathematical and physical sciences} \textbf{252}~(1271), 561--569 (1959).

\bibitem{kim2015crystal}
H.~Y. Kim, J.~Fu, H.~Tobe, J.~I. Kim, S.~Miyazaki, Crystal structure, transformation strain, and superelastic property of Ti--Nb--Zr and Ti--Nb--Ta alloys. \emph{Shape memory and Superelasticity} \textbf{1}~(2), 107--116 (2015).

\bibitem{hao2006effect}
Y.~Hao, S.~Li, S.~Sun, R.~Yang, Effect of Zr and Sn on Young's modulus and superelasticity of Ti--Nb-based alloys. \emph{Materials Science and Engineering: A} \textbf{441}~(1-2), 112--118 (2006).

\bibitem{obbard2011effect}
E.~Obbard, \emph{et~al.}, The effect of oxygen on $\alpha"$ martensite and superelasticity in Ti--24Nb--4Zr--8Sn. \emph{Acta Materialia} \textbf{59}~(1), 112--125 (2011).

\bibitem{chou2019oxygen}
K.~Chou, E.~A. Marquis, Oxygen effects on $\omega$ and $\alpha$ phase transformations in a metastable $\beta$ Ti--Nb alloy. \emph{Acta Materialia} \textbf{181}, 367--376 (2019).

\bibitem{nakai2009effect}
M.~Nakai, M.~Niinomi, T.~Akahori, H.~Tsutsumi, M.~Ogawa, Effect of oxygen content on microstructure and mechanical properties of biomedical Ti-29Nb-13Ta-4.6 Zr alloy under solutionized and aged conditions. \emph{Materials transactions} \textbf{50}~(12), 2716--2720 (2009).

\bibitem{wang2021roles}
J.~Wang, W.~Xiao, L.~Ren, Y.~Fu, C.~Ma, The roles of oxygen content on microstructural transformation, mechanical properties and corrosion resistance of Ti-Nb-based biomedical alloys with different $\beta$ stabilities. \emph{Materials Characterization} \textbf{176}, 111122 (2021).

\bibitem{tahara2016role}
M.~Tahara, T.~Inamura, H.~Y. Kim, S.~Miyazaki, H.~Hosoda, Role of oxygen atoms in $\alpha"$ martensite of Ti-20 at.\% Nb alloy. \emph{Scripta Materialia} \textbf{112}, 15--18 (2016).

\bibitem{mishin2021machine}
Y.~Mishin, Machine-learning interatomic potentials for materials science. \emph{Acta Materialia} \textbf{214}, 116980 (2021).

\bibitem{anstine2023machine}
D.~M. Anstine, O.~Isayev, Machine learning interatomic potentials and long-range physics. \emph{The Journal of Physical Chemistry A} \textbf{127}~(11), 2417--2431 (2023).

\bibitem{mortazavi2023atomistic}
B.~Mortazavi, X.~Zhuang, T.~Rabczuk, A.~V. Shapeev, Atomistic modeling of the mechanical properties: the rise of machine learning interatomic potentials. \emph{Materials horizons} \textbf{10}~(6), 1956--1968 (2023).

\bibitem{freitas2022machine}
R.~Freitas, Y.~Cao, Machine-learning potentials for crystal defects. \emph{MRS Communications} \textbf{12}~(5), 510--520 (2022).

\bibitem{liu2023three}
H.~Liu, \emph{et~al.}, Three-dimensional shape and stress field of a deformation twin in magnesium. \emph{Acta Materialia} \textbf{250}, 118845 (2023).

\bibitem{ledbetter1999habit}
H.~Ledbetter, M.~L. Dunn, Habit planes, inclusion theory, and twins. \emph{Materials Science and Engineering: A} \textbf{273}, 222--225 (1999).

\bibitem{lee1990elastic}
J.~Lee, M.~Yoo, Elastic strain energy of deformation twinning in tetragonal crystals. \emph{Metallurgical Transactions A} \textbf{21}~(9), 2521--2530 (1990).

\bibitem{yoo1991deformation}
M.~Yoo, J.~Lee, Deformation twinning in hcp metals and alloys. \emph{Philosophical Magazine A} \textbf{63}~(5), 987--1000 (1991).

\bibitem{yang2021332}
Y.~Yang, \emph{et~al.}, \{332\}$\langle$113$\rangle$ Twinning transfer behavior and its effect on the twin shape in a beta-type Ti-23.1 Nb-2.0 Zr-1.0 O alloy. \emph{Journal of Materials Science \& Technology} \textbf{91}, 58--66 (2021).

\bibitem{kwasniak2022polymorphic}
P.~Kwasniak, F.~Sun, S.~Mantri, R.~Banerjee, F.~Prima, Polymorphic nature of $\{$332$\}$< 113> twinning mode in BCC alloys. \emph{Materials Research Letters} \textbf{10}~(5), 334--342 (2022).

\bibitem{plimpton1995fast}
S.~Plimpton, Fast Parallel Algorithms for Short-Range Molecular Dynamics. \emph{J. Comput. Phys.} \textbf{117}~(1), 1--19 (1995), \doi{10.1006/jcph.1995.1039}.

\bibitem{thompson2022lammps}
A.~P. Thompson, \emph{et~al.}, {{LAMMPS}} - a Flexible Simulation Tool for Particle-Based Materials Modeling at the Atomic, Meso, and Continuum Scales. \emph{Comput. Phys. Commun.} \textbf{271}, 108171 (2022), \doi{10.1016/j.cpc.2021.108171}.

\bibitem{melchionna1993hoover}
S.~Melchionna, G.~Ciccotti, B.~Lee~Holian, Hoover NPT dynamics for systems varying in shape and size. \emph{Molecular Physics} \textbf{78}~(3), 533--544 (1993).

\bibitem{shinoda2004rapid}
W.~Shinoda, M.~Shiga, M.~Mikami, Rapid estimation of elastic constants by molecular dynamics simulation under constant stress. \emph{Physical Review B} \textbf{69}~(13), 134103 (2004).

\bibitem{bussi2007canonical}
G.~Bussi, D.~Donadio, M.~Parrinello, Canonical sampling through velocity rescaling. \emph{The Journal of chemical physics} \textbf{126}~(1) (2007).

\bibitem{ray1984statistical}
J.~R. Ray, A.~Rahman, Statistical ensembles and molecular dynamics studies of anisotropic solids. \emph{The Journal of chemical physics} \textbf{80}~(9), 4423--4428 (1984).

\bibitem{zhen2012deformation}
Y.~Zhen, C.~Chu, A deformation--fluctuation hybrid method for fast evaluation of elastic constants with many-body potentials. \emph{Computer Physics Communications} \textbf{183}~(2), 261--265 (2012).

\bibitem{nye1985physical}
J.~F. Nye, \emph{Physical properties of crystals: their representation by tensors and matrices} (Oxford university press) (1985).

\bibitem{larsen2016robust}
P.~M. Larsen, S.~Schmidt, J.~Schi{\o}tz, Robust structural identification via polyhedral template matching. \emph{Modelling and Simulation in Materials Science and Engineering} \textbf{24}~(5), 055007 (2016).

\bibitem{yang2020plastic}
Y.~Yang, P.~Castany, Y.~Hao, T.~Gloriant, Plastic deformation via hierarchical nano-sized martensitic twinning in the metastable $\beta$ Ti-24Nb-4Zr-8Sn alloy. \emph{Acta Materialia} \textbf{194}, 27--39 (2020).

\bibitem{gao2025manipulating}
Y.~Gao, \emph{et~al.}, Manipulating TWIP/TRIP via oxygen-doping to synergistically enhance strength and ductility of metastable beta titanium alloys. \emph{Journal of Materials Science \& Technology} \textbf{215}, 58--70 (2025).

\bibitem{zhan2015deformation}
H.~Zhan, W.~Zeng, G.~Wang, D.~Kent, M.~Dargusch, On the deformation mechanisms and strain rate sensitivity of a metastable $\beta$ Ti--Nb alloy. \emph{Scripta Materialia} \textbf{107}, 34--37 (2015).

\bibitem{bertrand2016deformation}
E.~Bertrand, P.~Castany, Y.~Yang, E.~Menou, T.~Gloriant, Deformation twinning in the full-$\alpha"$ martensitic Ti--25Ta--20Nb shape memory alloy. \emph{Acta Materialia} \textbf{105}, 94--103 (2016).

\end{thebibliography}
\bibliographystyle{sciencemag}

%
%
%
%
%
%


\section*{Acknowledgments}

\paragraph*{Funding:}
G.C. and L.Q. gratefully thank the funding support from National Science Foundation, United States, grant \#CMMI-2121866. D.V.P. and Y.Z. appreciate the financial support from National Science Foundation, grant \#2346524 and \#2515460.The calculations were performed by using the Extreme Science and Engineering Discovery Environment (XSEDE) Stampede2 at the TACC through allocation TG-MR190035. 
\paragraph*{Author contributions:}
G.C. conceived the idea and designed simulations, processed and analyzed the
data and wrote the manuscript. D.V.P. performed experiments, analyzed the experimental data. Y.Z. supervised the experiments and L.Q. supervised the project. Y.Z. and L.Q. acquired funding and resources. All authors revised the manuscript.
\paragraph*{Competing interests:}
There are no competing interests to declare.
\paragraph*{Data and materials availability:}
All key data refer to theories and conclusions are present in the paper and/or the Supplementary Materials.


\end{document}